\documentclass[prx,twocolumn,showpacs,superscriptaddress,preprintnumbers,amssymb]{revtex4-2}

\usepackage{bm}
\usepackage[dvipdfmx]{graphicx}
\usepackage{amsmath}
\usepackage{amssymb}
\usepackage{amsthm}
\usepackage{braket}
\usepackage{siunitx}
\usepackage{color}
\usepackage[utf8]{inputenc}
\usepackage{hyperref}
\usepackage{here}

\usepackage{tikz}
\usepackage{tikz-cd}
\usetikzlibrary{arrows}
\usetikzlibrary{intersections}
\usetikzlibrary{shapes.geometric}
\usetikzlibrary{decorations.pathmorphing, patterns,shapes}
\usetikzlibrary{decorations.markings}

\usepackage[most]{tcolorbox}

\hypersetup{ 
setpagesize=false,
 bookmarksnumbered=true,%
 bookmarksopen=true,%
 colorlinks=true,%
 linkcolor=blue,
 citecolor=blue,
 urlcolor=blue
}

\theoremstyle{plain}

\newtheorem*{thm*}{Theorem}
\theoremstyle{definition}

\usepackage{physics}
\usepackage{mathtools}

\newcommand{\secref}[1]{Sec.\,\ref{#1}}
\newcommand{\appref}[1]{Appendix.\,\ref{#1}}
\newcommand{\eqnref}[1]{Eq.\,\eqref{#1}}

\newcommand{\figref}[1]{Fig.\,\ref{#1}}
\newcommand{\tabref}[1]{Tab.\,\ref{#1}}

\begin{document}
\title{Coherent information for CSS codes under decoherence}
\author{Ryotaro Niwa}
\email{ryotaro.niwa@phys.s.u-tokyo.ac.jp}
\affiliation{Department of Physics, The University of Tokyo, 7-3-1 Hongo, Bunkyo-ku, Tokyo 113-0033, Japan}

\author{Jong Yeon Lee}\email{jongyeon@illinois.edu}
\affiliation{Department of Physics, The University of California, Berkeley, CA 94618, USA}
\affiliation{Department of Physics, University of Illinois at Urbana-Champaign, Urbana, Illinois 61801, USA}

\date{\today}

\begin{abstract}
Stabilizer codes lie at the heart of modern quantum-error-correcting codes (QECC). Of particular importance is a class called Calderbank-Shor-Steane (CSS) codes, which includes many important examples such as toric codes, color codes, and fractons. Recent studies have revealed that the decoding transition for these QECCs could be intrinsically captured by calculating information-theoretic quantities from the mixed state. Here we perform a simple analytic calculation of the coherent information for general CSS codes under local incoherent Pauli errors via diagonalization of the density matrices and mapping to classical statistical mechanical (SM) models. Our result establishes a rigorous connection between the decoding transition of the quantum code and the phase transition in the random classical SM model. 
It is also directly confirmed for CSS codes that exact error correction is possible if and only if the maximum-likelihood (ML) decoder always succeeds in the thermodynamic limit. Thus, the fundamental threshold is saturated by the optimal decoder.
 
\end{abstract}

\maketitle

\section{Introduction}
Quantum information encoded in a physical system is fragile against noise from the environment. 
Thus, one needs to devise a clever way to recover information from the decohered mixed state, a process called quantum error correction. Stabilizer codes~\cite{gottesman1997stabilizer} lie at the heart of such quantum-error-correcting schemes. Logical qubits are stored in the common eigenspace of mutually commuting operators called \emph{stabilizers}, and errors are detected by syndrome measurements. 
A special class of stabilizer codes with only $Z$-type and $X$-type stabilizers are called Calderbank-Shor-Steane (CSS) codes~\cite{CSS1, CSS2}. This class 
includes many paradigmatic examples of \textit{topological} quantum-error-correcting-codes (QECC) such as toric codes \cite{Kitaev_2003}, color codes~\cite{PhysRevLett.97.180501, Kubica_2015}, or fractons~\cite{PhysRevA.83.042330, PhysRevB.94.235157}.

Topological codes have many important properties. Perhaps the most remarkable one is that there exists an error-correcting decoder with success probability 1, provided that the system size $n$ is large ($n\,{\to}\,\infty$) and the physical error rate is under some finite threshold $p_{c}$ \cite{Dennis_2002,Katzgraber_2009}. It is widely accepted that this ``decoding transition'' is closely tied to the phase transition of the associated random-bond models along the Nishimori line \cite{Nishimori1981, Nishimori1986, PhysRevB.63.104422, Wang_2003, Ohzeki_2009, Kubica_2018, Chubb_2021}. Traditional arguments, however, were based on the success-fail transition of a specific decoder and did not deal directly with the mixed state after decoherence.

Recent work \cite{ lee2022symmetry, fan2023diagnostics, Lee_2023, bao2023mixedstate, chen2023separability, chen2023symmetryenforced, wang2023intrinsic, guo2023twodimensional, sang2023mixedstate, colmenarez2023accurate, PhysRevResearch.6.043258, su2024tapestry, lyons2024understanding,
lee2024exact,
li2024replicatopologicalorderquantum,
chen2024unconventionaltopologicalmixedstatetransition} on decoherence-induced phase transitions (DIPT) provides a new perspective to this problem. In particular, it was shown in \cite{fan2023diagnostics, Lee_2023} that various information theoretic quantities of the decohered toric code under local bit/phase-flip channels could be mapped to some physical quantities in the underlying classical statistical mechanical (SM) model, exhibiting transition behavior at a certain critical error-rate.  Without specifying any particular decoder, this transition could be interpreted as an \textit{intrinsic} property of the mixed-state topological phase.

A generalization of the aforementioned work to various stabilizer codes was later considered in \cite{PhysRevResearch.6.043258, su2024tapestry, lyons2024understanding}, where Rényi entropic quantities are calculated for the decohered stabilizer codes. Notably, Ref.~\cite{PhysRevResearch.6.043258} employs a replica limit to derive an analytic expression for the coherent information, with the limit rigorously justified using Carlson's theorem.
However, calculating Rényi entropic quantities and taking the Replica limit to recover the physical quantities involves some subtleties and complications. 
A recent paper resolved this problem for the specific example of the 2D toric code, by exactly calculating \textit{coherent information} via diagonalization of the mixed state density matrices \cite{lee2024exact}. This made the connection between the decoding transition of the 2D toric code and the criticality of the 2D RBIM more concrete. 

In light of these developments, here we perform an analytic calculation of the coherent information for general CSS codes under local incoherent Pauli errors, via diagonalization of the mixed state density matrices and mapping to classical SM models. Our result rigorously establishes a connection between the decoding transition of the quantum code and the phase transition in the underlying random classical SM model. It also directly confirms that exact error correction is possible if and only if the maximum-likelihood (ML) decoder always succeeds in the thermodynamic limit~\footnote{Depending on the perspective, it is often called a maximum-entropy decoder as well.}. Thus, the fundamental error threshold is saturated by the optimal decoder. Furthermore, we show how the decoding threshold based on \textit{relative entropy} upper-bounds the fundamental threshold in the thermodynamic limit, and emphasize the importance of comparing these two points carefully.  
We also provide a systematic way to construct the underlying SM model based on the classical codes used to construct the CSS code.

The rest of the paper is organized as follows. In section \ref{Background}, we give a brief review of CSS codes, including concepts such as \textit{CSS chain complex}. In section \ref{setup}, we review on some facts about coherent information and describe our noise model. In section \ref{Results}, we present our coherent information calculation and the mapping to random classical SM models. Finally, in section \ref{Discussion}, we summarize our results and point out some future directions.

\section{Background}
\label{Background}

CSS codes are stabilizer codes that have either $X$-type stabilizers $\hat{A}_{s}$ (a tensor product of Pauli-$X$) or $Z$-type stabilizers $\hat{B}_{p}$ (a tensor product of Pauli-$Z$). They are mutually commuting 
\begin{equation}\label{commuting}
    [\hat{A}_{s},  \hat{B}_{p}] = 0,
\end{equation}
and the logical space is defined by the condition
\begin{equation}\label{ground}
    \hat{A}_{s} \ket{\psi} = \ket{\psi}, \hat{B}_{p}  \ket{\psi} = \ket{\psi} \, (\forall s,p).
\end{equation}
These stabilizers generate the \textit{stabilizer group} $\cal S$. Pauli operators commuting with all elements of $\cal S$ forms a normalizer group ${\cal N}_{\cal P}({\cal S})$~\footnote{It is equivalent to the centralizer for the stabilizer group.}, and logical operators are elements of ${\cal N}_{\cal P}({\cal S})\,{\setminus}\,{\cal S}$.

CSS codes can be constructed from a pair of classical linear codes. Given bit-strings $\bm{x}$ of length $n$ with $x_i\,{\in}\,\mathbb{F}_2\,{=}\,\{0,1\}$, a classical code $\cal C$ is completely defined by its parity check matrix $H$ such that any valid codeword $\bm{x}\,{\in}\,\mathbb{F}_2^n$ satisfies $H \bm{x}\,{=}\,\bm{0}$ (Note that the operation is done in $\mathbb{F}_{2}$). The number of logical bits is $k\,{=}\,\dim(\ker H)$, and the code distance is $d\,{=}\,\min_{x \in \ker H} \sum_i x_i$. 

Let $\mathcal{C}_{z}, \mathcal{C}_{x}$ be two codes with same $n$, whose parity check matrices $H_{z}, H_{x}$ satisfy
\begin{equation}\label{twolevel}
    H_{x}^{\vphantom{T} } H_{z}^{T} = 0.
\end{equation}
The \textit{stabilizer matrix} can then be constructed as 
\begin{equation} \label{eq:stabilizerS}
    S= \mqty[H_{z} & 0 \\
    0 & H_{x}].
\end{equation}
Each row of $S$ corresponds to the binary representation of a stabilizer, where the first and second $n$ columns correspond to $Z$ and $X$ operators respectively. \eqnref{twolevel} implies that all stabilizers constructed this way commute, giving rise to a CSS code.

Given $S$ in \eqnref{eq:stabilizerS}, the associated logical operators can be found by defining the \textit{normalizer matrix} $N$:
\begin{equation}\label{normalizer}
    N = \mqty[O & G_{z}\\
    G_{x} & O],
\end{equation}
where $G_a$ is an $n\times k_a$ generator matrix that maps a bitstring of length $k_a$ into a codeword in ${\cal C}_a$, satisfying $H_a G_a\,{=}\,\bm{0}$. Each column of $N$ corresponds to the binary representation of an operator consisting of Pauli $Z$ and $X$ operators in a similar manner. By performing \textit{symplectic-Gram-Schmidt orthogonalization procedure} (SGSOP) \cite{Wilde_2009}, we can obtain $k$ pairs of anti-commuting normalizers, which are precisely the logical operators of the code (See Appendix \ref{apA}). From this procedure, we can confirm for arbitrary CSS codes that all logical-$X/Z$ operators can be set as $X/Z$-type, respectively.

A useful concept in analyzing CSS codes is the \textit{CSS chain complex} \cite{kubica2018ungauging}. By regarding \eqnref{twolevel} as the \textit{exactness} of boundary maps in $\mathbb{Z}_{2}$ homology, Z-type stabilizers, qubits, X-type stabilizers can be thought of as forming a length-3 chain complex. Here, $H_{z}^{T}$ and $H_{x}$ are boundary operators, and $H_{x}^{T}$ and $H_{z}$ are coboundary operators.

\begin{figure}[t]
    \centering
    \includegraphics[width=0.8\columnwidth]{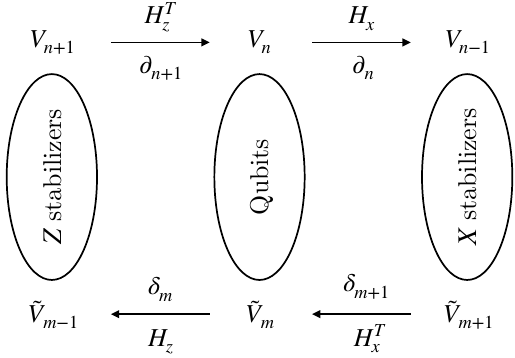}
    \caption{{\bf CSS chain complex}. $Z$-type stabilizers ($V_{n+1}$), qubits ($V_{n}$), and $X$-type stabilizers ($V_{n-1}$) can be thought of as a chain complex of length-3. When we can embed the chain complex in some manifold where $V_m$ is a set of $m$-cells, parity check matrices and their transposes play the role of boundary operators $\partial$. One can also consider a dual chain complex consisting of $\tilde{V}_m$s and coboundary operators $\delta$.}
    \label{CSSchain}
\end{figure}

Various CSS codes, including the toric code and the color code, have SM models originating from their underlying classical codes. Given a classical code $\cal C$ with a parity check matrix $H$, 
A classical bit $\sigma_{i}$ is connected with check $C_{a}$, if and only if $\sigma_{i}$ is included in $C_{a}$. The transpose of the parity check matrix $H^T$ can be interpreted as the biadjancecy matrix that maps a check into bits included in the check. 
Then, one can always construct a Hamiltonian for classical spins $\{ \sigma_i= \pm 1 \}$ as follows:
\begin{align} \label{clgr}
    H_{\mathrm{cl}} = -\sum_{a} C_{a} &= -\sum_{a} \prod_{b=1}^n \sigma_{b}^{H_{ab}},
\end{align}
where the summation $a$ is over checks, i.e., rows of $H$. The groundstates of $H_{\mathrm{cl}}$ correspond to valid codewords by $\bm{x}\,{=}\,(\bm{\sigma}\,{+}\,\bm{I})/2$, where  $\bm{I}\,{=}\,(1,1,\cdots 1)$. Conversely, given any classical Hamiltonian for Ising spins, one can construct a classical code. 
If there is an error that flips individual spin with a random probability, a process of going from one groundstate configuration to another can be understood by an SM model taking \eqnref{clgr} at a finite temperature.
Therefore, we employ this perspective throughout the paper and use the terms ``classical codes'' and ``classical SM models'' interchangeably.

\section{Setup}
\label{setup}

Let us now describe our setup. To probe the transition in the maximum amount of decodable information in a system (code) $Q$, we use the \textit{coherent information} as our diagnostic. 
Let $k$ denote the number of logical qubits in $Q$. 
To proceed, we introduce two systems, $R$ and $R'$ with $\dim R = \dim R' = 2^k$, which are maximally entangled. Here $R$ represents the reference qubits, while $R'$ (eventually) corresponds to the logical space of the code. 
The system $R'$ is then combined with an ancillary system $A$ through an encoding map $\Xi$ to form the code $Q\,{=}\,R'A$, which encodes $k$ logical qubits redundantly.
Finally, the decoherence channel $\cal E$ is applied on the system $Q$, which can be modeled as a unitary operator ${\cal U}_{\cal E}$ applied between $Q$ and the environment $E$~\footnote{Note that, given ${\cal E}$, coherent information is independent of the choice of $U_{\cal E}$~\cite{Schumacher_1996}.}. The coherent information $I_c(R,Q;{\cal E})$ is defined as
\begin{equation}
    I_c(R,Q;{\cal E}) = S(\rho_{Q})-S(\rho_{QR}).
    \begin{tikzpicture}[scale=0.65,baseline={(current bounding box.center)}]
    	\draw[thick] (-0.8, 3.2)node[left]{\scriptsize$R$} -- (-0.8, 0) -- (0.5, 0) -- (0.5,1);
    	\draw[thick] (0.5,0)node[below]{\scriptsize$R'$} -- (0.5,1);
    	\draw[thick] (1.1,0)node[below]{\scriptsize$A\vphantom{'}$} -- (1.1,1);
    	\draw[line width=0.6mm] (0.8,1) -- (0.75,3.2)node[left]{\scriptsize$Q$};
    	\draw[thick] (2.3,0.5) -- (2.3, 3.2)node[left]{\scriptsize$E$};
    	\filldraw [fill=white, thick, rounded corners=0.2cm] (0.2, 1.5) rectangle ++(2.55, 1) node [midway] {\scriptsize${\cal U}_{\cal E}$};
    	\filldraw [fill=white, thick, rounded corners=0.2cm] (0.2, 0.5) rectangle ++(1.2, 0.5) node [midway] {\scriptsize$\Xi$};
    	\draw[thick] (1.8,0.5) -- (2.8, 0.5) -- (2.3, 0) -- (1.8, 0.5);
    \end{tikzpicture}
\end{equation} 
Before the action of $\mathcal{E}$ on system Q, which is equivalent to take ${\cal U}_{\cal E}\,{=}\,1$, $\rho_{QR}$ is a pure state with maximal entanglement between $Q$ and $R$ and $I_{c}(\rho_{0,Q}) = k \log 2$. Using coherent information, the exact quantum error correction condition can be rephrased as the following Nielsen-Schumacher condition~\cite{Schumacher_1996}:
\begin{equation}\label{NS}
    I_{c}(\rho_{Q}) = S(\rho_{R}) = k \log 2,
\end{equation}
which means that the quantum information between the reference $R$ and $Q$ does not leak to the environment. 
By the data processing inequality~\cite{Schumacher_1996}, $I_{c}$ is monotonically decreasing with respect to the error and no recovery operation $\mathcal{R}$ exists once this condition is violated. Thus, a critical rate $p_{c}$ at which $I_{c}$ becomes strictly smaller than $k \log 2$ gives a rigorous upper bound for the error threshold. The subadditivity of the entanglement gives the lower bound that $-S(R) = - k \log 2 \leq  I_c$.

We start from the maximally mixed logical subspace of an arbitrary CSS stabilizer code with $k$ logical qubits.
\begin{equation}
    \rho_{0,Q} = \frac{1}{2^k}\prod_{s}\qty(\frac{1+\hat{A}_{s}}{2})\prod_{p}\qty(\frac{1+\hat{B}_{p}}{2})
\end{equation}
We then apply the local bit/phase-flip channel
\begin{equation}
    \mathcal{E}_{a,i}[\rho_{0,Q}] = (1-p_{a})\rho_{0,Q}+p_{a}\sigma_{i}^{a}\rho_{0,Q}\sigma_{i}^{a} \quad (a=x,z), 
\end{equation}
with $\mathcal{E}_{a} = \prod_{i} \mathcal{E}_{a,i}$, $\mathcal{E} =  \mathcal{E}_{z} \circ \mathcal{E}_{x}$. 
A generic density matrix under Pauli noises can be written as
\begin{equation}
    \rho_{Q} = \mathcal{E}[\rho_{0,Q}] = \hspace{-3pt} \sum_{E_x, E_z} \hspace{-3pt} P(E_x,E_z) Z_{E_z} X_{E_x}\rho_{0,Q}X_{E_x} Z_{E_z}.
\end{equation}
where $E_a$ is an $a$-type error-chain, $X_E = \prod_{i \in E} \sigma^x_i$, and $Z_E = \prod_{i \in E} \sigma^z_i$. Here $P(E_x,E_z) = P(E_x) P(E_z)$, where
\begin{align}
P_{a}(E)&=p_{a}^{|E|}(1-p_{a})^{n-|E|}.
\end{align}
Although we first address this independent $X$ and $Z$ noise case for simplicity, our formalism allows us to calculate coherent information against a depolarization channel with correlated $X$ and $Z$ noises, see \secref{sec:corrXZ} and appendix \ref{Ynoise}.

\section{Results}
\label{Results}

\subsection{Coherent information via diagonalization}

Given a CSS code, one can find $k$ independent logical operators $\{ K_i \}$, then the complete orthonormal basis of the physical Hilbert space is labeled in terms of parity of $\{ A_s \}$, $\{ B_p \}$, and $\{ K_i \}$ as $|\vb{a}, \vb{b}, \vb{k} \rangle$. For example, one can consider either $\{ \bar{\sigma}^x_i \}_{i=1}^k$ or $\{ \bar{\sigma}^z_i \}_{i=1}^k$ as a set of $k$ independent $X$-type or $Z$-type logical operators.

To exactly calculate coherent information, we first express $\rho_{0,Q}$ in a diagonal form 
\begin{equation}
    \rho_{0,Q} = \frac{1}{2^{k}}\sum_{\vb{k}}\ket{\bm{0}_x, \bm{0}_z, \vb{k}}\bra{\bm{0}_x, \bm{0}_z, \vb{k}},
\end{equation}
where $\bm{0}_{a}\,{=}\,(0,0,\cdots ,0)\,{\in}\,\mathbb{F}_{2}^{m_a}$ represents the parity of the independent $a$-type stabilizers, and $\vb{k}\in \mathbb{F}_{2}^{k}$ represents the logical state. Note that $m_x + m_z = n-k$. 
The density matrix after applying the bit-flip channel is
\begin{equation}
    {\cal E}_x[\rho_{0,Q}] = \frac{1}{2^{k}}\sum_{\vb{b}}P_{\vb{b}}\sum_{\vb{k}}\ket{\bm{0}_x,\vb{b}, \vb{k}}\bra{\bm{0}_x,\vb{b}, \vb{k}},
\end{equation}
where $P_{\vb{b}}$ represents the probability that the parity for $Z$-type stabilizers become $\vb{b} \in \mathbb{F}_{2}^{m_{z}}$. Note that 
\begin{align} \label{eq:Pb}
    P_{\vb{b}} = \sum_{E | H_z E = \vb{b}}  p_{x}^{|E|}(1-p_{x})^{n-|E|}
\end{align}
where $E \in \mathbb{F}_2^n$. Also, even if the error-chain $X_E$ amounts to some logical operators $ \bar{\sigma}_{i}^{a}\, (a=x,z)$, the ``maximally-mixedness'' of $\rho_{0,Q}$ ensures 
\begin{equation} \label{eq:mixedness} \bar{\sigma}_{i}^{a}\sum_{\vb{k}}\ket{\vb{k}}\bra{\vb{k}}\bar{\sigma}_{i}^{a}=\sum_{\vb{k}}\ket{\vb{k}}\bra{\vb{k}}.
\end{equation}
Similarly, applying the phase-flip error, we get
\begin{equation}\label{Q}
    \rho_Q = {\cal E}[\rho_{0,Q}]=\frac{1}{2^{k}}\sum_{\vb{a}, \vb{b}, \vb{k}}P_{\vb{a}}P_{\vb{b}}\ket{\vb{a}, \vb{b},\vb{k}}\bra{\vb{a}, \vb{b},\vb{k}},
\end{equation}
where $P_{\vb{a}}\,{=}\,\sum_{E|H_x E =\bm{a}} p_z^{|E|} (1\,{-}\,p_z)^{n-|E|}$ is the probability that the $X$-type stabilizer parity become $\vb{a} \in \mathbb{F}_{2}^{m_{x}}$. The above expression completes our diagonalization of $ \rho_{Q}$, which is independent of the choice of logical basis $|\bm{k} \rangle$. 

We next diagonalize $\rho_{QR}$. We start from $\rho_{0, QR}$, which is maximally entangled with the reference qubits:
\begin{align}
    &\rho_{0, QR} =\prod_{i, a}\qty(\frac{1+\bar{\sigma}_{i}^{a}\otimes\tau^{a}_{i}}{2})\rho_{0,Q}\otimes I_{R}\nonumber\\
&=\frac{1}{4^k}\prod_{i=1}^{k}\qty[\sum_{n_x, n_z=0}^1  (\bar{\sigma}_{i}^{x}\otimes\tau^{x}_{i})^{n_x} ( \bar{\sigma}_{i}^{z}\otimes\tau^{z}_{i})^{n_z} ]\rho_{0,Q}.
\end{align}
Here, $\tau_{i}^{a}$ represents Pauli matrices for the reference qubits, and we have omitted the identity operator $I_{R}$ for the reference system. Using the property in \eqnref{eq:mixedness}, after applying ${\cal E}_x \circ {\cal E}_z$, we obtain the form 
\begin{align}\label{QR}
    \rho_{QR}=\hspace{-11pt} \sum_{\vb{k}, \vb{a}, \vb{b}, \vb{k}_{x}, \vb{k}_{z}, } \hspace{-11pt} 
 P_{\vb{a}, \vb{k}_{x}} P_{\vb{b}, \vb{k}_{z}}\qty[\prod_{i=1}^{k} \rho_{i}^{(\vb{k}_z, \vb{k}_x)}] \ket{\vb{a}, \vb{b}, \vb{k}} \bra{\vb{a}, \vb{b}, \vb{k}},
\end{align}
where $P_{\vb{a}, \vb{k}_{x}}$ represent the probability that the error-chain induces the parity change of $\vb{a}$ in the stabilizers and $\vb{k}_x$ in the logical-$X$ operators (or put differently, applying the logical $Z$-operator $\prod_i (\bar{\sigma}_i^z)^{k_{x,i}}$), and similarly for $P_{\vb{b}, \vb{k}_{z}}$.  
The key observation here is that 
\begin{align}\label{ortho}
\rho_{i}^{(\vb{k}_z, \vb{k}_x)}&= \sum_{n_a=0}^1 \frac{1}{4}  (e^{i\pi k_{x,i}} \bar{\sigma}_{i}^{x}\otimes\tau^{x}_{i})^{n_x} (e^{i\pi k_{z,i}} \bar{\sigma}_{i}^{z}\otimes\tau^{z}_{i})^{n_z}, 
\end{align}
due to the nontrivial commutation relation $[\bar{\sigma}_i^z, \bar{\sigma}_j^x ] = 2 i \delta_{ij} \bar{\sigma}^y_i$.
It is straightforward to show that $\rho_{i}^2=\rho_{i}$, thus each $\rho_i$ is a projector. With this representation, define 
\begin{align}
    {\cal P}(\vb{k}_z, \vb{k}_x) := \prod_{i=1}^k \rho_{i}^{(\vb{k}_z, \vb{k}_x)}.
\end{align}
Then, ${\cal P}(\vb{k}_z, \vb{k}_x)$ is a projection operator onto orthogonal subspaces labeled by $(\vb{k}_z, \vb{k}_x)$ since ${\cal P}(\vb{k}_z, \vb{k}_x) {\cal P}(\vb{k}'_z, \vb{k}'_x) = {\cal P}(\vb{k}_z, \vb{k}_x) \delta_{\vb{k}_{z}^{\prime}-\vb{k}_{z}}\delta_{\vb{k}_{x}^{\prime}-\vb{k}_{x}}$.
Thus, (\ref{QR}) completes our diagonalization of $ \rho_{QR}$. Moreover, comparing (\ref{Q}-\ref{ortho}) gives
\begin{equation}\label{summation}
    P_{\vb{a}}=\sum_{\vb{k}_x}P_{\vb{a},\vb{k}_{x}}, \quad    P_{\vb{b}}=\sum_{\vb{k}_z}P_{\vb{b},\vb{k}_{z}}.
\end{equation}
Altogether, from (\ref{Q}), (\ref{QR}), (\ref{summation}) we get 
\begin{align}\label{coherentinformation}
    I_{c}=k \log 2 &+\sum_{\vb{a}, \vb{k}_x}P_{\vb{a}, \vb{k}_x}\log \frac{P_{\vb{a}, \vb{k}_x}}{\sum_{\vb{k}_x^{ \prime}}P_{\vb{a},\vb{k}_x^{ \prime}}}\nonumber \\
    &+\sum_{\vb{b}, \vb{k}_z}P_{\vb{b}, \vb{k}_z}\log \frac{P_{\vb{b}, \vb{k}_z}}{\sum_{\vb{k}_z^{\prime}}P_{\vb{b},\vb{k}_z^{ \prime}}}.
\end{align}
We note that a similar idea was discussed in \cite{DiVincenzo_1998} for a single logical qubit. The diagonalization under independent $Y$ noise is also straightforward, although the probability distribution $P_{\vb{a}, \vb{b}, \vb{k}_x, \vb{k}_z}$ for a change in stabilizers and logical labels would not factor into two parts as in the above cases. See appendix \ref{Ynoise} for Y noise. 

\subsection{Mapping to random classical SM models: independent bit and phase flip noises}

Next, let us ask how to obtain the exact expression of $P_{\vb{a}, \vb{k}_x}$ and $P_{\vb{b}, \vb{k}_z}$ from the original quantum code. Various studies have already shown that these probabilities can be mapped to the partition function of some classical SM model \cite{Dennis_2002, Katzgraber_2009, Chubb_2021, Song_2022}, in the context of ML decoding. Here, we follow the traditional idea, but with a special emphasis on the role of parity check matrices and symmetries.

\begin{figure}[t]
    \centering
    \includegraphics[width=0.55\columnwidth]{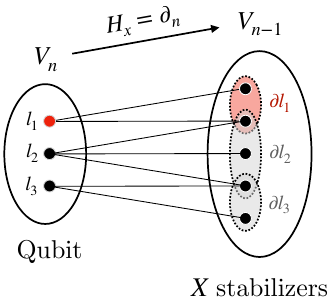}
    \caption{
    {\bf Mapping to classical SM models for $X$ errors}. For bit-flip (Pauli-$X$) errors, virtual classical spins are placed on $(n\,{-}\,1)$-cells in $V_{n-1}$ ($X$ stabilizers). For each qubit, we can identify a corresponding set of $X$-type stabilizers supported on that qubit. The product of the classical spins in such a set gives each term in the SM model. 
    Here, an $X$ error chain $\vb{E}_{\vb{b}}$ is denoted by red dots in $V_n$. If the term originates from red qubits (red circle in $V_{n-1}$), it is antiferromagnetic with a positive sign in the Hamiltonian. If not (gray circle in $V_{n-1}$), it is ferromagnetic with a negative sign. Notice that the relation between the qubits and the $X$-type stabilizers is captured by the parity check matrix of the underlying classical linear code $\mathcal{C}_x$.}
    \label{mapping}
\end{figure}

Without loss of generality, consider a bit-flip ($X$ error) chain.
If it incurs a parity change in $Z$-stabilizers by $\vb{b}$, there are many error chains $\vb{E} \in \mathbb{F}_2^n$ satisfying $\vb{b} = H_z \vb{E}$.  
Pick a representative chain $\vb{E}_0$ satisfying $H_z \vb{E}_0 = \vb{b}$. 
The group of error chains $\{\vb{E}\}$ satisfying $\vb{b} = H_z \vb{E}$ can be denoted as $\vb{E}_0 \,{+}\, \ker H_z$, which is isomorphic to $\ker H_z$. Notably, $\ker H_z$ can be decomposed as
\begin{align} \label{eq:decomposition}
    \ker H_z \cong \frac{\ker(H_z)}{\Im(H_x^T)} \oplus \Im(H_x^T)
\end{align}
where the first component corresponds to the logical $X$ operator, thus the parity change in logical-$Z$ eigenvalues~\cite{bombin2013introduction, bravyi2013homological}.

Fixing the logical parity change to be $\vb{k}_z$, we aim to parametrize $\Im(H_x^T)$. The isomorphism theorem implies
\begin{align}
    \Im(H_x^T)\simeq  \mathbb{F}^{m'_x}_2 / \ker(H_x^T)
\end{align}
where $m'_x$ is the number of columns of $H_x^T$ (or rows of $H_x$), i.e., the number of parity check operators that are not necessarily independent. Defining $D_x\,{:=}\,\dim\big[\ker(H_x^T) \big]$, for a fixed $(\vb{b},\vb{k}_z)$, the following family parametrizes $\ker(H_z)$ with the degeneracy of $2^{D_x}$:
\begin{align}
    \{ \vb{E}^{\vb{k}_z}_{\vb{b}} + H_x^T \bm{\sigma} \,|\, \bm{\sigma} \in \mathbb{F}_2^{m_x'} \}.
\end{align} 
where $\vb{E}^{\vb{k}_z}_{\vb{b}}$ corresponds to some particular error-chain which specifies $(\vb{b}, \vb{k}_z)$. 
For example, in the 2D toric code $D_{x}\,{=}\,1$. Combining this parametrization with \eqnref{eq:Pb}, it follows that 
\begin{equation}
    P_{\vb{b}, \vb{k}_z}  =  \frac{(1-p_{x})^{n}}{2^{D_x}} \sum_{\bm{\sigma}}\qty(\frac{p_{x}}{1-p_{x}})^{|\vb{E}^{\vb{k}_z}_{\vb{b}} + H_x^T \bm{\sigma}|}.
\end{equation}
The summation over $\bm{\sigma}$ is equivalent to the summation over $X$-type stabilizers $\{A_s \}$, which leads us to place a virtual  binary variable $\sigma_s\,{=}\,0, \, 1$  on every stabilizer $s$ as in \figref{mapping}. $\bm{\sigma}$ lives in the Boolean space $\mathbb{F}_2^{m_x'}$ whose dimension is larger than the number of independent $X$ stabilizers. 
Then, one can show that
\begin{align}
    \big|  \vb{E}^{\vb{k}_z}_{\vb{b}}  + H_x^T \bm{\sigma}  \big| &= \frac{1}{2} \Big( n -  \sum_{l=1}^n  e^{i\pi( \vb{E}_{\vb{b}}^{\vb{k}_z} + H_x^T \bm{\sigma})_l } \Big) \nonumber \\
    &= \frac{n}{2}  -  \frac{1}{2} \sum_{l=1}^n  e^{i\pi E_{\vb{b},l}^{\vb{k}_z}} \hspace{-10pt} \prod_{s | H_{x,sl} \neq 0} \hspace{-5pt}  e^{i \pi \sigma_s}   
\end{align} 
where the summation is over all edge qubits labeled by $l$ as in \figref{mapping}. With $e^{i \pi \sigma_s}\,{=}\,\pm 1$, this is the sum of Ising interactions among spins that share the same edge $l$. In what follows, we use $\sigma_s$ as if it is an Ising variable $e^{i \pi \sigma_s}$.

From the perspective introduced in \figref{CSSchain}, the condition $H_{x,sl} \neq 0$ is equivalent to $s \in \partial l$ where $\partial$ represents the boundary map within the geometry defined by the parity check operator. 
Finally, the classical model has the symmetry group isomorphic to $\ker(H_x^T)$. For any $\bm{a},{\in}\,\ker(H_x^T)$, $\sigma_s \mapsto a_s \sigma_s$ leaves the Hamiltonian invariant. 
Therefore, $P_{\vb{b}, \vb{k}_z}$ corresponds to a partition function of the classical SM model with quenched-disorder $\mathcal{Z}_{x}(\vb{b}, \vb{k}_{z})$ where
\begin{align}
    \label{partitionfunction}
    P_{\vb{b}, \vb{k}_z}  &=  \frac{\mathcal{Z}_{x}(\vb{b}, \vb{k}_{z})}{2^{D_x} (2 \cosh \beta_x)^{n}} \nonumber \\
    \mathcal{Z}_{x}(\vb{b}, \vb{k}_{z}) &=  \sum_{\bm{\sigma}}  e^{\beta_x \sum_l (-1)^{E^{\vb{k}_z}_{\vb{b},l}} \prod_{s \in \partial l}   \sigma_s},
\end{align}
where $\beta_x=-\frac{1}{2}\mathrm{log}\frac{p_{x}}{1-p_{x}}$, which is equivalent to a so-called Nishimori condition~\cite{Nishimori1981, Nishimori1986}.

Similarly for a $Z$ error chain incurring a parity change in  syndrome observation by $\vb{a}$ and logical $X$ operator by $\vb{k}_x$, it can be parametrized as
\begin{align}
    \{ \vb{E}^{\vb{k}_x}_{\vb{a}} + H_z^T \bm{\tau} \,|\, \bm{\sigma} \in \mathbb{F}_2^{m_z'} \}.
\end{align} 
where $\vb{E}^{\vb{k}_x}_{\vb{a}}$ corresponds to some particular error-chain which specifies $(\vb{a}, \vb{k}_x)$, and $m_z'$ is the number of rows of $H_z$. With this identification, $P_{\vb{a}, \vb{k}_x}$ is given as
\begin{align}
    \label{partitionfunction}
    P_{\vb{a}, \vb{k}_x}  &=  \frac{\mathcal{Z}_{z}(\vb{a}, \vb{k}_{x})}{2^{D_z} (2 \cosh \beta_z)^{n}} \nonumber \\
    \mathcal{Z}_{z}(\vb{a}, \vb{k}_{x}) &=  \sum_{\bm{\tau}}  e^{\beta_z \sum_l (-1)^{E^{\vb{k}_x}_{\vb{a},l}} \prod_{p \in \delta l}   \tau_p},
\end{align}
where $\vb{E}^{\vb{k}_x}_{\vb{a}}$ is an error chain inducing the parity change in logical-$X$ operator by $\vb{k}_x$ with $H_x \vb{E}^{\vb{k}_x}_{\vb{a}} = \vb{a}$, classical spins $\bm{\tau}\,{=}\,\{\tau_p\}$ are located at the center of $Z$-type stabilizers $\{B_p \}$, and the coboundary operator $\delta\,{:=}\,H_z$.

As remarked in \eqnref{eq:decomposition}, $\vb{k}_{x,z}$ is the element of  the quotient group $\ker(H_x)/\Im(H_z^T) \cong \ker(H_z)/\Im(H_x^T)$ (as $H_x H_z^T\,{=}\,0$). With a proper embedding in manifold ${\cal M}$, it corresponds to the $n$-th homology group ${\cal H}_n({\cal M}) = \ker( \partial_n )/\Im( \partial_{n+1} )$~\footnote{In a more general situation, this is captured by relative homology group.}.
For example, in the 2D toric code, $n\,{=}\,1$ and ${\cal H}_1({\cal M})$ would be the first homology group characterizing nontrivial cycles of ${\cal M}$.

Finally, two SM models for ${\cal Z}_x$ and ${\cal Z}_z$ are Krammers-Wannier dual to each other at the frustration-free limit, which is a natural consequence of the condition for the CSS code $H_x^{\vphantom{T}} H_z^T\,{=}\,0$ as elaborated in \appref{app:KW}. 
This duality is exact when it comes to Renyi-($m\,{\geq}\,2$) quantities~\cite{su2024tapestry, lyons2024understanding} where corresponding SM models are frustration-free. However, for the correct von-Neumann information theoretic quantities, the duality between random ensembles does not hold exactly anymore.

\subsection{Correlated bit and phase flip noises} \label{sec:corrXZ}

In the previous section, we have studied SM models corresponding to eigenvalues of the density matrix in \eqnref{QR}. As remarked, due to the anti-commutativity of $Y$ error, even when we introduce $Y$ error the only change is that the eigenvalue $P_{\vb{a}, \vb{k}_{x}} P_{\vb{b}, \vb{k}_{z}}$ loses its factored structure, becoming $P_{\vb{a}, \vb{b}, \vb{k}_{x}, \vb{k}_{z}}$. More generally, we can consider a depolarization channel
\begin{align}
    {\cal E}_i[\rho] = (1-\tilde{p}) \rho + \sum \tilde{p}_a \sigma^a_i \rho \sigma_i^a
\end{align}
where $\tilde{p}\,{=}\,\tilde{p}_x \,{+}\, \tilde{p}_y \,{+}\, \tilde{p}_z$. Independent bit and phase flip noise channels can be expressed as a single depolarization channel with the following identification:
\begin{align}
    \tilde{p}_x &= p_x(1-p_z) \nonumber \\
    \tilde{p}_z &= p_z(1-p_x) \nonumber \\
    \tilde{p}_y &= p_x p_z.
\end{align}
Thus, any deviation from this implies that bit/phase flip errors are correlated. In such a case, we obtain that
\begin{align} \label{eq:generalP}
    P_{\vb{a}, \vb{b}, \vb{k}_{x}, \vb{k}_{z}} &=  (1-\tilde{p})^{n} \sum_{ \tilde{E} } \prod_{i=x,y,z} \qty(\frac{\tilde{p}_i}{1-\tilde{p}} )^{|\tilde{E}_i|} 
\end{align}
where $\tilde{E}\,{=}\,\sum_i \tilde{E}_i$. Given $(\vb{a}, \vb{b}, \vb{k}_x, \vb{k}_z)$, we can parameterized all possible $X$ and $Z$ error strings consistent with this as
\begin{align}
    E_x &= E^{\vb{k}_z}_{\vb{b}} + H_x^T \bm{\sigma} \in \mathbb{F}_2^n \nonumber \\
    E_z &= E^{\vb{k}_x}_{\vb{a}} + H_z^T \bm{\tau} \in \mathbb{F}_2^n,
\end{align}
where ($E^{\vb{k}_z}_{\vb{b}},E^{\vb{k}_x}_{\vb{a}}$) is a representative error string for this syndrome observation.
With this, $\tilde{E}_i$ in \eqnref{eq:generalP} becomes
\begin{align}
    \tilde{E}_x &= E_{x} (\bm{1}- E_{z}) \nonumber \\
    \tilde{E}_z &= E_{z} (\bm{1} - E_{x}) \nonumber \\
    \tilde{E}_y &= E_{x} E_{z}.
\end{align}
where $\bm{1} = (1,1,...,1) \in \mathbb{F}_2^n$ and the product is element-wise multiplication between two $n$-dimensional vectors. 
The product between $E_x$ and $E_z$ would generate interactions between $\bm{\sigma}$ and $\bm{\tau}$. Therefore, we obtain that
\begin{align} \label{eq:SMcorr}
& P_{\vb{a}, \vb{b}, \vb{k}_{x}, \vb{k}_{z}}  \propto {\cal Z}(\vb{a}, \vb{b}, \vb{k}_{x}, \vb{k}_{z}) := \sum_{\bm{\sigma}, \bm{\tau}} e^{-\sum_{i,l} H_{i,l}} \nonumber \\
    &H_{y, l} = \qty( -\frac{\tilde{\beta}_y}{2} + \frac{\tilde{\beta}_x}{2} + \frac{\tilde{\beta}_z}{2}  )  (-1)^{E^{\vb{k}_z}_{\vb{b},l}+ E^{\vb{k}_x}_{\vb{a},l}} \prod_{s \in \partial l}   \sigma_s \prod_{p \in \delta l}   \tau_p  \nonumber \\
    &H_{x, l} = \qty( \frac{\tilde{\beta}_x}{2} - \frac{\tilde{\beta}_z}{2} + \frac{\tilde{\beta}_y}{2}  )  (-1)^{E^{\vb{k}_z}_{\vb{b},l}} \prod_{s \in \partial l}   \sigma_s \nonumber \\
    &H_{z, l} = \qty( \frac{\tilde{\beta}_z}{2} - \frac{\tilde{\beta}_x}{2} + \frac{\tilde{\beta}_y}{2}  )  (-1)^{E^{\vb{k}_x}_{\vb{a},l}} \prod_{p \in \delta l}   \tau_p
\end{align}
where $e^{-2 \tilde{\beta}_i} = \frac{\tilde{p}_i}{1 - \tilde{p}}$. 
The SM model for correlated $X$ and $Z$ errors in \eqnref{eq:SMcorr} has a symmetry group given as a direct sum of symmetries of SM$_x$ and SM$_z$, i.e., $\ker(H_x^T)$ and $\ker(H_z^T)$. For self-dual CSS codes with ${\cal C}_x = {\cal C}_z$, it can exhibit a global symmetry further enhanced by the duality.

\subsection{Exact error correction}
Now that we have exactly calculated coherent information, we can discuss the condition to perform exact quantum error correction. Comparing the Nielsen-Schumacher condition in  \eqnref{NS} with \eqnref{coherentinformation}, we notice that 
\begin{equation}\label{condition}
    \sum_{\vb{a}, \vb{k}_{x}}P_{\vb{a}, \vb{k}_{x}}\mathrm{log}\frac{P_{\vb{a}, \vb{k}_{x}}}{\sum_{\vb{k}_{x}^{ \prime}}P_{\vb{a}, \vb{k}_{x}^{ \prime}}} = \sum_{\vb{b}, \vb{k}_{z}}P_{\vb{b}, \vb{k}_{z}}\mathrm{log}\frac{P_{\vb{b}, \vb{k}_{z}}}{\sum_{\vb{k}_{z}^{ \prime}}P_{\vb{b},\vb{k}_{z}^{ \prime}}}=0
\end{equation}
is necessary and sufficient.  Let us compare this condition with the exact QEC condition for the optimal decoder. Given a syndrome observation $(\vb{a},\vb{b})$, if we calculate the partition function for the associated random bond model, we obtain the relative probability among different values of $(\vb{k}_x, \vb{k}_z)$. The optimal decoder generates the output based on this probability distribution; accordingly, when we have an error $(\vb{a},\vb{b}, \vb{k}_x, \vb{k}_z)$, the probability of success  for the optimal decoder is given as
\begin{align} \label{eq:condition1}
   P^{\textrm{suc.}}_{\vb{a},\vb{b}, \vb{k}_x, \vb{k}_z} = \frac{ P_{\vb{a},\vb{b}, \vb{k}_x, \vb{k}_z} }{ \sum_{\vb{k}'_x,\vb{k}'_z }P_{\vb{a},\vb{b}, \vb{k}'_x, \vb{k}'_z} }.
\end{align}
The total success probability is then given as
\begin{align}
\overline{ P^{\textrm{suc.}} } = \sum_{\vb{a}, \vb{b}, \vb{k}_x,\vb{k}_z} P_{\vb{a},\vb{b}, \vb{k}_x, \vb{k}_z} \cdot P^{\textrm{suc.}}_{\vb{a},\vb{b}, \vb{k}_x, \vb{k}_z}
\end{align}
since the probability of getting an error labeled by ${\vb{a},\vb{b}, \vb{k}_x, \vb{k}_z}$ is $P_{\vb{a},\vb{b}, \vb{k}_x, \vb{k}_z}$. We remark that the lower bound of this success probability is determined by the condition in \eqnref{eq:condition1}. By the Jensen's inequality, 
\begin{align}
    \overline{ \log P^\textrm{suc.}  } \leq \log \overline{ P^\textrm{suc.} }
\end{align}
where 
\begin{align}
    \overline{ \log P^\textrm{suc.}  } &=  \sum_{\vb{a}, \vb{k}_{x}}P_{\vb{a}, \vb{k}_{x}}\mathrm{log}\frac{P_{\vb{a}, \vb{k}_{x}}}{\sum_{\vb{k}_{x}^{ \prime}}P_{\vb{a}, \vb{k}_{x}^{ \prime}}} \nonumber \\
    & \quad + \sum_{\vb{b}, \vb{k}_{z}}P_{\vb{b}, \vb{k}_{z}}\mathrm{log}\frac{P_{\vb{b}, \vb{k}_{z}}}{\sum_{\vb{k}_{z}^{ \prime}}P_{\vb{b},\vb{k}_{z}^{ \prime}}}
\end{align}
in the case where $P_{\vb{a},\vb{b}, \vb{k}_x, \vb{k}_z}$ factorizes into $X$ and $Z$ parts. Therefore, one can establish that 
\begin{align}
    e^{I_c - k \log 2} \leq \overline{ P^{\textrm{suc.}} } \leq 1.
\end{align}
Note that the above success probability $P^{suc.}$ of a random sampling method provides a lower bound for the maximum likelihood method where for given $(\vb{a},\vb{b})$, we deterministically choose $(\vb{a},\vb{b}, \vb{k}_x, \vb{k}_z)$ with the largest probability (or smallest free energy):
\begin{equation}
    \overline{ P^{\textrm{suc.}} } = \hspace{-8pt} \sum_{\vb{a}, \vb{b}, \vb{k}_x,\vb{k}_z} \hspace{-8pt} P_{\vb{a},\vb{b}, \vb{k}_x, \vb{k}_z} \cdot P^{\textrm{suc.}}_{\vb{a},\vb{b}, \vb{k}_x, \vb{k}_z} \leq \sum_{\vb{a}, \vb{b}} \max_{\vb{k}_x, \vb{k}_z} P_{\vb{a}, \vb{b}, \vb{k}_x, \vb{k}_z}
\end{equation}
This is because $\sum p^2\,{\leq}\,\max p$. See also appendix \ref{app:optimal} for the ML decoder. Therefore, the condition that $I_c = k \log 2$ gives a sufficient condition for the optimal decoder to succeed always. As the coherent information is the upper bound on the maximum amount of decodable information, this implies that the thresholds for the coherent information and optimal decoder agree:
\begin{align}
    p_\textrm{th}^\textrm{opt} = p_\textrm{th}^\textrm{coh.}.
\end{align}
Thus, while the free energy provides an upper bound on the optimal decoder's performance~\cite{lee2024exact}, the coherent information provides the lower bound.

\subsection{Connection to Nishimori Physics}
One interesting feature here is that the eigenvalue of the density matrix $\rho_{QR}$ has the form $P_{\vb{a},\vb{b},\vb{k}_x,\vb{k}_z} \propto \mathcal{Z} (\vb{a}, \vb{k}_x) \mathcal{Z} (\vb{b}, \vb{k}_z)$ under the independent Pauli noises. As a result, the transition behavior of the coherent information is tied to the transition behavior in the associated random disordered classical model along the \textit{Nishimori line}~\cite{Nishimori1981, Nishimori1986}. A disordered classical model is on the Nishimori line if the inverse temperature is equal to $\beta_p$, a value associated with the distribution of the disorder (random coupling) in the system. One of the consequences is that the probability of having a certain disorder configuration in the ensemble is proportional to its partition function, which enables various analytical calculations. 

In the study of disordered classical models, disorder averaged squared (Edward-Anderson) order parameter or free energy of domain-wall insertion is often used to detect the phase transition. This coincides with a conventional way to derive the threshold for the ML decoder~\cite{Dennis_2002}; since the relative probability of different logical sectors is given as $e^{\Delta F}$ where $\Delta F$ is the free energy cost of inserting domain walls for the associated configuration, the disorder averaged value $\langle \Delta F \rangle$ has been used to identify the transition. Furthermore, the disorder averaged domain wall free energy is equivalent to the \textit{relative entropy} between different logical sectors $D(\rho||\rho^{\prime})\,{\propto}\,\langle \Delta F \rangle$, an information-theoretic metric to characterize decodability, see appendix \ref{app:relative}. 
Thus, one can define the transition point $p_{\textrm{th}}^{\textrm{rel}}$ as the critical point above which relative entropy stops diverging.

However, is $p_{\textrm{th}}^{\textrm{coh}} = p_{\textrm{th}}^{\textrm{rel}}$? Relative entropy may in general fail to characterize the decoding transition. It is possible that $I_c<k\log 2$ even though $D(\rho||\rho^{\prime})\,{\rightarrow}\,\infty$. To rigorously prove that $p_\textrm{th}^{\textrm{coh}}$ agrees with  $p_{\textrm{th}}^{\textrm{rel}}$, one needs to be careful. At least, it can be shown~\cite{Kubica_2018} that in the thermodynamic limit,
\begin{equation}\label{eq:necessary}
I_c=k \log 2\,\, \implies\,\, D(\rho||\rho^{\prime}) = \infty,
\end{equation} 
as elaborated in appendix \ref{app:optimal}. This implies 
\begin{equation}
     p^{\mathrm{dec}}_{\textrm{th}} \leq p_{\textrm{th}}^{\textrm{coh}} \leq p_{\textrm{th}}^{\textrm{rel}},
\end{equation}
where $p^{\mathrm{dec}}_{\textrm{th}}$ is the error threshold for an arbitrary decoder. Note that in the literature, $p_{\textrm{th}}^{\textrm{rel}}$ is also identified as the Nishimori critical point for random bond models. 
This proves that the Nishimori critical point is indeed a rigorous upper bound for the threshold of an arbitrary decoder. However, a logical possibility that $p_\textrm{th}^\textrm{coh.}$ may be smaller than $p_{\textrm{th}}^{\textrm{rel}}$ for some general infinite family of codes is not excluded, although they coincide for the specific example of the 2D Toric Code case~\cite{Dennis_2002, Wang_2003, lee2024exact}.

\begin{table*}
\caption{\label{tab:table3}The associated classical SM models of some well-known CSS codes. 
Random-interaction versions of these models along the Nishimori line have been discussed in the context of ML decoding~\cite{Kubica_2018, Katzgraber_2009, Ohzeki_2009, Song_2022}.}
\begin{ruledtabular}
\begin{tabular}{ccc}
 CSS stabilizer code & $\mathcal{C}_{x}$ & $\mathcal{C}_{z}$ \\ \hline
2D Toric Code \cite{Dennis_2002} &2D Ising model &2D Ising model\\
3D Toric Code &3D Ising model & 3D plaquette gauge model\\
(4,8,8) 2D Color Code \cite{Katzgraber_2009, Ohzeki_2009} &2D Union-Jack Ising model & 2D Union-Jack Ising model\\
(6,6,6) 2D Color Code \cite{Katzgraber_2009, Ohzeki_2009} & 2D Triangle Ising model & 2D Triangle Ising model\\
3D Color Code \cite{Kubica_2018} & 4-body Ising model & 6-body Ising model\\
X-cube model \cite{Song_2022} & 3D plaquette Ising model & 3D anisotropically-coupled Ashkin-Teller model
\end{tabular}
\end{ruledtabular}
\end{table*}

\subsection{Examples}
To illustrate how the transition behavior of coherent information is tied to the phase transition in the associated random SM model, we provide a detailed description of the underlying classical code for some prominent CSS codes. Many topological codes have well-known random-interaction SM models dictated by their underlying classical codes, which have been discussed in the context of ML decoding or higher Renyi quantities~\cite{Dennis_2002, Ohzeki_2009, Kubica_2018, Chubb_2021, fan2023diagnostics, Lee_2023, lee2024exact, lyons2024understanding, su2024tapestry}. See \tabref{tab:table3} for a concise summary. Here, we derive them systematically with a special focus on symmetries and domain walls.

\subsubsection{Example: 2D Surface Code}

{\bf Setup.} The 2D Surface Code \cite{bravyi1998quantum, Dennis_2002} is defined on a 2D square lattice with open boundary conditions. It has two types of boundaries, \emph{smooth} (left and right) and \emph{rough} (top and bottom). 
While the smooth boundary is terminated with vertices connected by edges, the rough boundary is determined by the protruded edges.  
The stabilizers are defined at vertices and plaquettes as 
\begin{equation}
    \hat{A}_{v} = \prod_{i\in \partial v} X_{i}, \quad \hat{B}_{p} = \prod_{i \in \partial p} Z_{i}.
\end{equation}

{\bf Symmetry.} Unlike the toric code case, all stabilizers are independent: $\dim(\ker H_x^T)\,{=}\,\dim(\ker H_z^T)\,{=}\,0$. This reflects the fact that the underlying classical code has no symmetry; in the SM model, while the bulk part of the classical Hamiltonian is the 2D RBIM Ising model, the boundary part has single spin terms (Zeeman fields) that break the global spin-flip symmetry.  

\begin{figure}[!t]
\centering
\includegraphics[width=1\columnwidth]{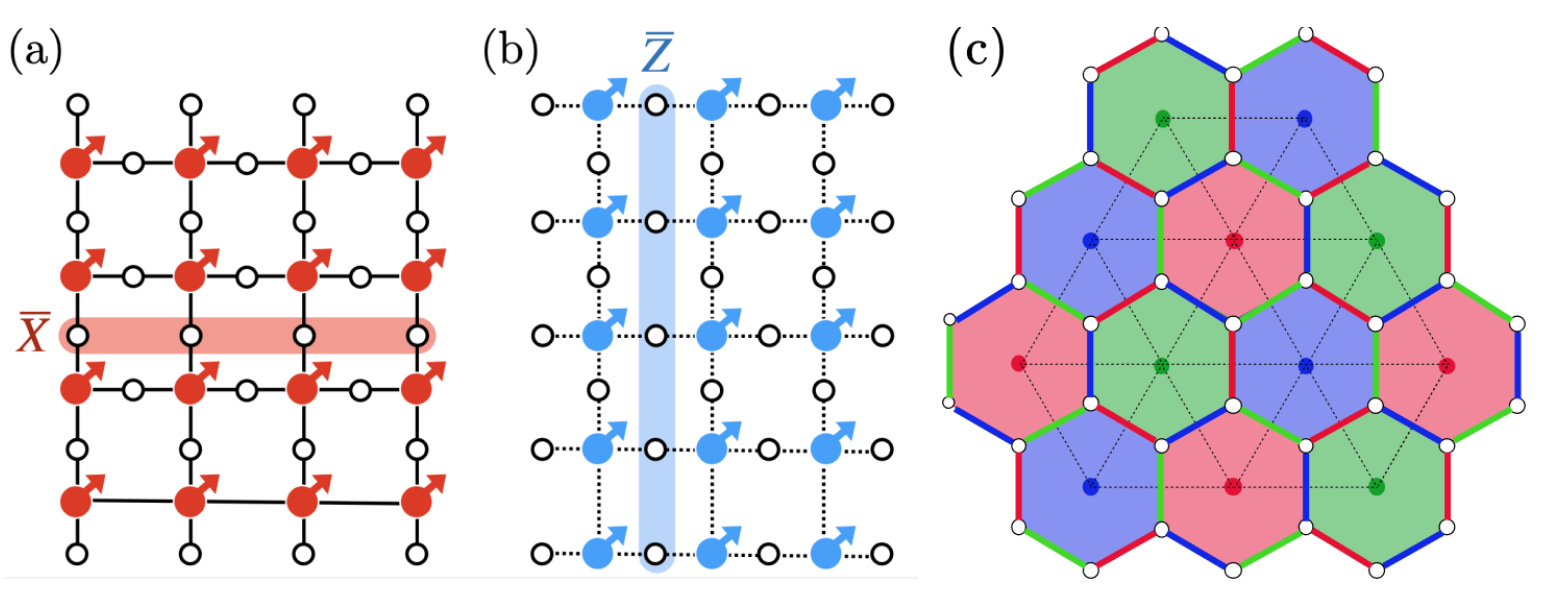}
\caption{\label{fig:2dmodel}
 {\bf 2D surface code for (a) SM$_x$ and (b) SM$_z$}. For $X$ error chain, SM$_x$ is defined for spins on vertices of the original lattice (solid line). The logical operator $\bar{X}$ connects the left and right boundaries. 
 For $Z$ error chain, SM$_z$ is defined for spins on plaquettes, or vertices of the dual lattice as drawn in the figure (dotted line). The logical operator $\bar{Z}$ connects the top and bottom boundaries. 
For SM$_x$ (SM$_z$), there are boundary Zeeman fields at the top and bottom (left and right) spins. 
{\bf (c) (6,6,6) 2D color code}. There are three different types of string operators transporting anyons with three different colors. 
However, the multiplication of two colors results in string operators with the other color, so only two of them are independent. For example, the product of the blue loop and the red loop is equal to the green loop multiplied by red plaquettes. The underlying classical SM model is the three-body Ising model defined on the dual lattice.}
\end{figure}

{\bf Logical space.} Following \eqnref{eq:decomposition}, the logical $X$ operator is the string operator which connects the two smooth boundaries as in \figref{fig:2dmodel}(a), and the logical $Z$ operator is the string operator which connects the two rough boundaries as in \figref{fig:2dmodel}(b). Note that a finite rectangle has a trivial homology group. 
However, when we have a rough boundary, $H_x$ maps an edge qubit at a rough boundary into a single vertex ($X$-stabilizer), which means that they do not have a proper 1-cell structure (since a normal edge has two vertices).
To equip a CSS chain complex with a proper manifold, one has to assume a virtual vertex ($X$-stabilizer) for each edge qubit in rough boundaries.
Then, we calculate topological property \emph{relative} to these virtual vertices since they do not exist in the CSS chain complex. 
With this understanding, $\ker H_x/\Im H_z^T$ corresponds to a relative homology group, where cycles are defined \emph{relative} to a set of virtual vertices placed at two rough boundaries. 
Let $S$ be a square and $A$ be a union of top and bottom edges. Then it will count equivalence classes for the paths $l$ with $\partial l \in A$ (\emph{relative cycle}) that cannot be contracted to the top or bottom edges.
Indeed, a nontrivial element of $\ker H_x/\Im H_z^T\,{=}\,{\cal H}_1(S,A;\mathbb{Z}_2)\,{=}\,\mathbb{Z}_2$ corresponds to a path between top and bottom boundaries, which cannot be deformed into a subset of $A$~\cite{Dennis_2002, Zhu_2022}.

{\bf SM$_{x,z}$.} To be concrete, consider the SM model for $X$ error chain as in \figref{fig:2dmodel}(a). There are random Zeeman fields at rough boundaries (top and bottom). However, the coherent information calculation still boils down to comparing partition functions with and without a domain wall across left and right, which corresponds to identifying whether logical operator $\overline{X}$ has been applied. 
As $(i)$ we are interested in the domain wall structure across horizontal direction while the boundary fields are at the top and bottom, and $(ii)$ boundary fields cannot affect the bulk critical physics, the transition point we obtain would be the same as the toric code example with explicit symmetry and periodic boundary condition. The SM model for $Z$ error chain behaves analogously, where spins are placed at the center of plaquettes (or vertices of the dual lattice) as in \figref{fig:2dmodel}(b).

\subsubsection{Example: 2D Color Code}

{\bf Setup.} The 2D Color Code~\cite{PhysRevLett.97.180501, Kubica_2015, landahl2011faulttolerant} is defined on a two-dimensional lattice with three-colorable tilling and periodic boundary conditions as in \figref{fig:2dmodel}(c).
In this three-colorable lattice, each face can be colored by red (R), green (G) or blue (B), with each vertex neighboring three faces with different colors. 
The qubits are placed on the vertices, and the stabilizers are defined on each face as 
\begin{equation}
    \hat{A}_{f} = \prod_{s\in \partial f} X_{s}, \quad \hat{B}_{f} = \prod_{s \in \partial f} Z_{s}.
\end{equation}
Because each face contains an even number of vertices and any neighboring faces overlap on an even number of vertices, these stabilizers commute.

{\bf Symmetry.}  Due to the periodic boundary condition and the three-colorable tiling, the product of all stabilizers on the faces of any two specific colors should be the identity.
\begin{equation}
    \prod_{f \in \bm{R},\bm{G}} \hat{A}_{f} = \prod_{f \in \bm{G},\bm{B}} \hat{A}_{f} = \prod_{f \in \bm{B},\bm{R}} \hat{A}_{f} = 1.
\end{equation}
This holds similarly for $Z$ stabilizers.  Taking 
\begin{equation*}
    \prod_{f \in \bm{R},\bm{G}} \hat{A}_{f}\cdot \prod_{f \in \bm{G},\bm{B}} \hat{A}_{f}=\prod_{f \in \bm{B},\bm{R}} \hat{A}_{f}
\end{equation*}
into account, we have $\dim(\ker H_x^T)\,{=}\,2$, meaning that there are two independent global symmetries for $\mathcal{C}_{x}$. Therefore, a global symmetry action for SM$_x$ can be labeled by a pair of colors.
The analogous logic follows for $\mathcal{C}_{z}$, giving $\dim(\ker H_z^T)\,{=}\,2$.

{\bf Logical space.} In the 2D color code, logical operators are again given by the loops around nontrivial cycles of the underlying lattice. More precisely, one can define a loop operator for each color along each nontrivial cycle.  
Without loss of generality, consider red plaquettes and assign red colors for edges connecting red plaquettes as in \figref{fig:2dmodel}(c). Then, a logical operator $\bar{X}_R$ is defined as the product of Pauli-$X$s on red edges along the non-trivial cycle of the torus. One can proceed analogously for the other two colors blue and green for each nontrivial cycle. However, it can be shown that the product of two logical $X$ operators for two colors is equivalent to the logical $X$ operator for the other color up to stabilizers. For example, $\bar{X}_R \bar{X}_G = \bar{X}_B$. 
Therefore, for each nontrivial cycle, there are two independent logical $X$ (or $Z$) operators; since there are two nontrivial cycles in $\mathbb{T}^2$, there are four logical qubits.

{\bf SM$_x$, SM$_z$.} To be concrete, consider the SM model for $X$ error chain in \figref{fig:2dmodel}(c). We place classical spins at the center of hexagonal plaquettes (vertices of the dual triangular lattice). As each vertex qubit is shared by three neighboring $X$-type stabilizers, the SM model is defined on the triangular dual lattice with random three-body interactions~\cite{Katzgraber_2009}. Note that three sublattices of the dual triangular lattice can have color labels inherited from the original plaquettes. 

The aforementioned global symmetries $\ker H_x^T$ correspond to sublattice Ising symmetries on ($R,G$), $(G,B)$, and $(B,R$) sublattices. Notice that only two of these three symmetries are independent. 
It is straightforward to verify that the domain wall for any symmetry labeled by two colors $(c,c')$ corresponds to the logical $X$ operator with color $c'' \neq c,c'$ along the domain wall.
The behavior of SM$_z$ against a $Z$ error is derived analogously.

\subsubsection{Example: 3D Toric Code}

{\bf Setup.}  The 3D Toric Code is defined on the 3D cubic lattice with periodic boundary conditions. The qubits are placed on the edges, and the stabilizers are defined on vertices $\{v\}$ and faces $\{f\}$: 
\begin{equation}
    \hat{A}_{v} = \prod_{e\in \delta v} X_{e}, \quad \hat{B}_{f} = \prod_{e \in \partial f} Z_{e}.
\end{equation}
If we have $n$ cubes, then we have $n$ vertices, $3n$ edges, and $3n$ faces. 

{\bf Symmetry.} The stabilizers have the following constraints: 
\begin{equation}
    \prod_{v} \hat{A}_{v} = 1, \quad \prod_{f\in S} \hat{B}_{f} =1,
\end{equation}
where $S$ corresponds to a closed surface. 
With $\dim(\ker H_x^T)=1$ and $\dim(\ker H_z^T)=n+2$, these constraints translate into a 0-form global symmetry in SM$_x$ and extensive numbers of 1-form symmetries in SM$_z$.

{\bf Logical space.} There are three logical qubits. For each direction $i$, there are logical $X$ and $Z$ operators; $\bar{X}_i$ is the product of $X$ operators on bonds along the $i$ direction across the perpendicular surface. $\bar{Z}_i$ is the product of $Z$ operators one the non-contractible loop wrapping the $\mathbb{T}^3$ along the $i$-direction.

{\bf SM$_x$.} There are $X$-type stabilizers $A_v$ for every vertex. Since two vertices share each edge qubit, the resulting SM model has spins on vertices with nearest-neighbor 2-body interactions; this is exactly the 3D RBIM. 
With the global Ising 0-form symmetry, the corresponding global domain wall is a $(d\,{-}\,1)$-dimensional object, which is a two-dimensional non-contractible surface, an element of ${\cal H}_2(\mathbb{T}^3; \mathbb{Z}_2) \simeq \mathbb{Z}_2^3$.
Accordingly, logical X operators are membrane operators along the $xy, yz, zx$ plane. Thus, changing the global frustration pattern $\vb{k}_{z}$ corresponds to flipping the interaction terms along the $\varepsilon_{ijk} \hat{k}$ direction across the $\hat{i}\hat{j}$ plane in the corresponding 3D RBIM.

{\bf SM$_z$.} There are $Z$-type stabilizers $B_f$ for every face. Since each edge qubit is shared by four faces, it consists of 4-body interactions. As there is an extensive number of closed surfaces, the constraint $\prod_{f \in S} B_f =1$ implies an extensive number of symmetries, which are 1-form symmetries.  
The resulting SM model is a 3D random plaquette gauge model (RPGM). 
With the 1-form symmetry, the corresponding global domain wall is a $(d\,{-}\,2)$-dimensional object, which is a one-dimensional non-contractible loop, an element of ${\cal H}_1(\mathbb{T}^3; \mathbb{Z}_2) \simeq \mathbb{Z}_2^3$~\footnote{As the symmetry is acting on the two-dimensional closed surface, the domain wall appears at the boundary of two-dimensional open surface, which is a closed one-dimensional loop.}
Thus, changing the parity $\vb{k}_{z}$  corresponds to favoring violated plaquette terms around the non-contractible loop in the 3D RPGM.

\begin{figure}[t]
    \centering
    \includegraphics[scale=0.3]{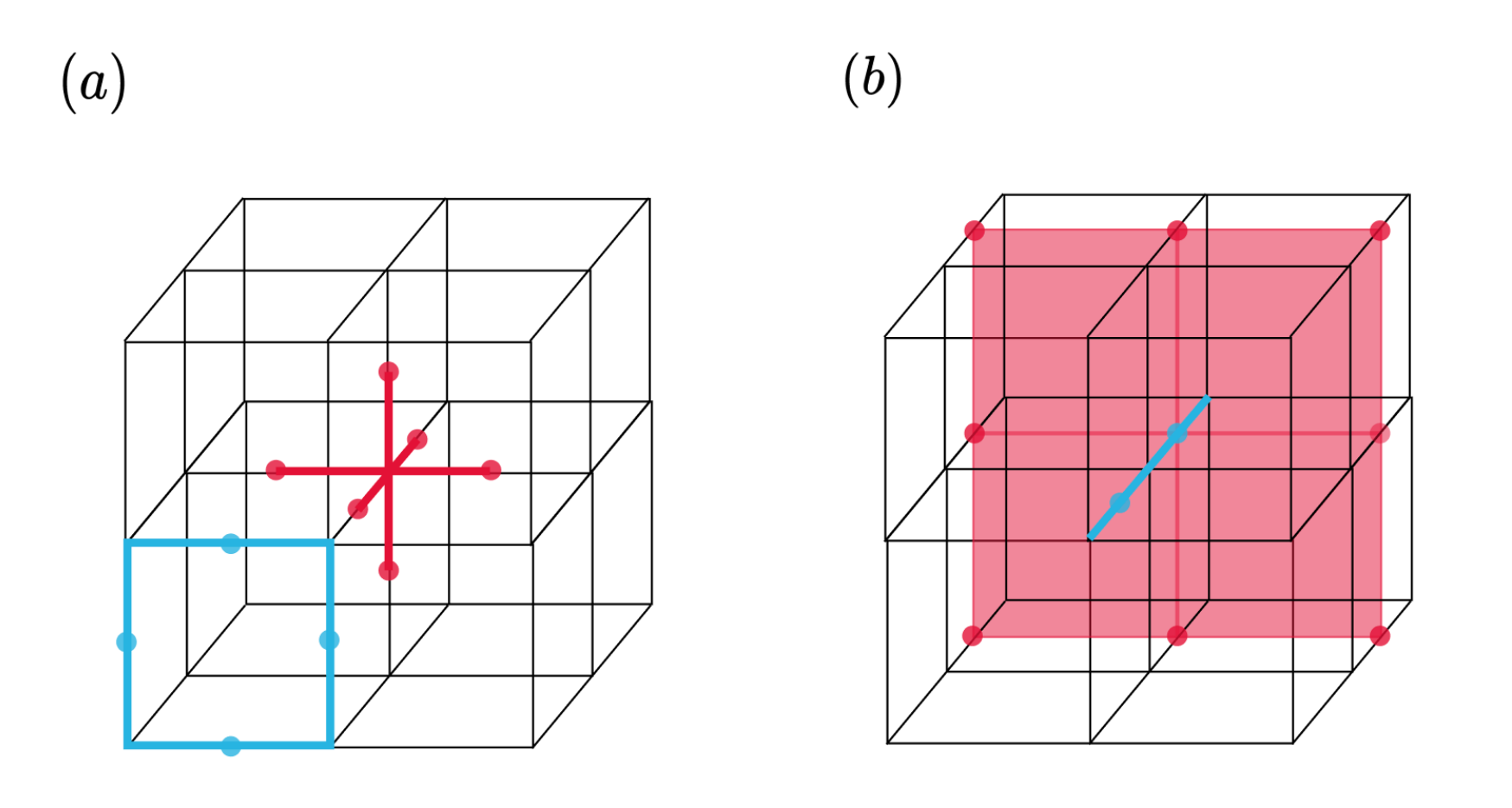}
    \caption{\textbf{3D toric code.} (a) Stabilizers defined on vertices and faces. Each edge qubit is shared by 2 vertex stabilizers, and 4 face stabilizers. (b) Logical $X$ operators are two-dimensional surfaces, while Logical $Z$ operators are string operators.}
\end{figure}

\subsubsection{Example: X-cube model}
{\bf Setup.} The X-cube model \cite{PhysRevB.94.235157, Shirley_2018} is defined on the 3D cubic lattice with periodic boundary conditions. The qubits are placed on the edges, and the stabilizers are defined on cubes $\{c\}$ and vertices $\{v\}$:
\begin{equation}
    \hat{A}_{c} = \prod_{e\in \partial c} X_{e}, \quad \hat{B}_{v}^{\mu} = \prod_{e \in \partial v, e\perp \mu} Z_{e},
\end{equation}
where $\mu=\hat{x}, \hat{y}, \hat{z}$ indicates that the edges $e$ lie in the plane perpendicular to the $\mu$ direction. Note that $\hat{B}_{v}^{x}\hat{B}_{v}^{y}\hat{B}_{v}^{z}=1$, thus $\hat{B}^\mu_v \hat{B}^\nu_v = |\epsilon_{\mu \nu \lambda}| \hat{B}^\lambda_v$.

{\bf Symmetry.} The stabilizers have the following constraints: 
\begin{equation}
    \prod_{c \in \Pi^*} \hat{A}_{c} = 1, \quad \prod_{v\in \Pi_\mu} \hat{B}^\mu_{v} =1,
\end{equation}
where $\Pi^*$ is a closed plane in the dual lattice and $\Pi_\mu$ is a closed plane whose normal vector is along $\hat{\mu}$ in the original lattice. Therefore, we have an extensive number of subsystem symmetries acting on $3L$ planes for both SM$_x$ and SM$_z$. 

{\bf Logical space.} Considering $L\,{\times}\,L\,{\times}\,L$ cubic lattice, there are $6L\,{-}\,3$ logical qubits~\cite{Shirley_2018}, $2L\,{-}\,1$ for each direction $i\,{\in}\,\{x,y,z\}$. Without loss of generality, consider the $\hat{z}$-direction. 
Here, each logical $X$ operator is defined as the product of $X$ operators on the straight lines along the $z$-direction. While there are $L^2$ such straight lines, $L^2-2L+1$ of them are equivalent to each other up to stabilizers; therefore, there are $2L-1$ independent $X$ operators along the direction $i$.
To define a logical $Z$ operator, pick a specific $xy$-plane of $z$-directional bonds (so its $z$-coordinate will be half-integer). 
Then, for each straight $x$ or $y$ line, we can define a logical $Z$ operator as the product of $Z$s on the $z$-directional bond along the line. There are $2L$ such lines, but the product of all of them is identity, so there are $2L-1$ independent logical $Z$ operators, matching the number of independent logical $X$ operators. Note that a different choice of plane with a different $z$-coordinate is equivalent up to $Z$-type stabilizers.

\begin{figure}[t]
    \centering
    \includegraphics[scale=0.4]{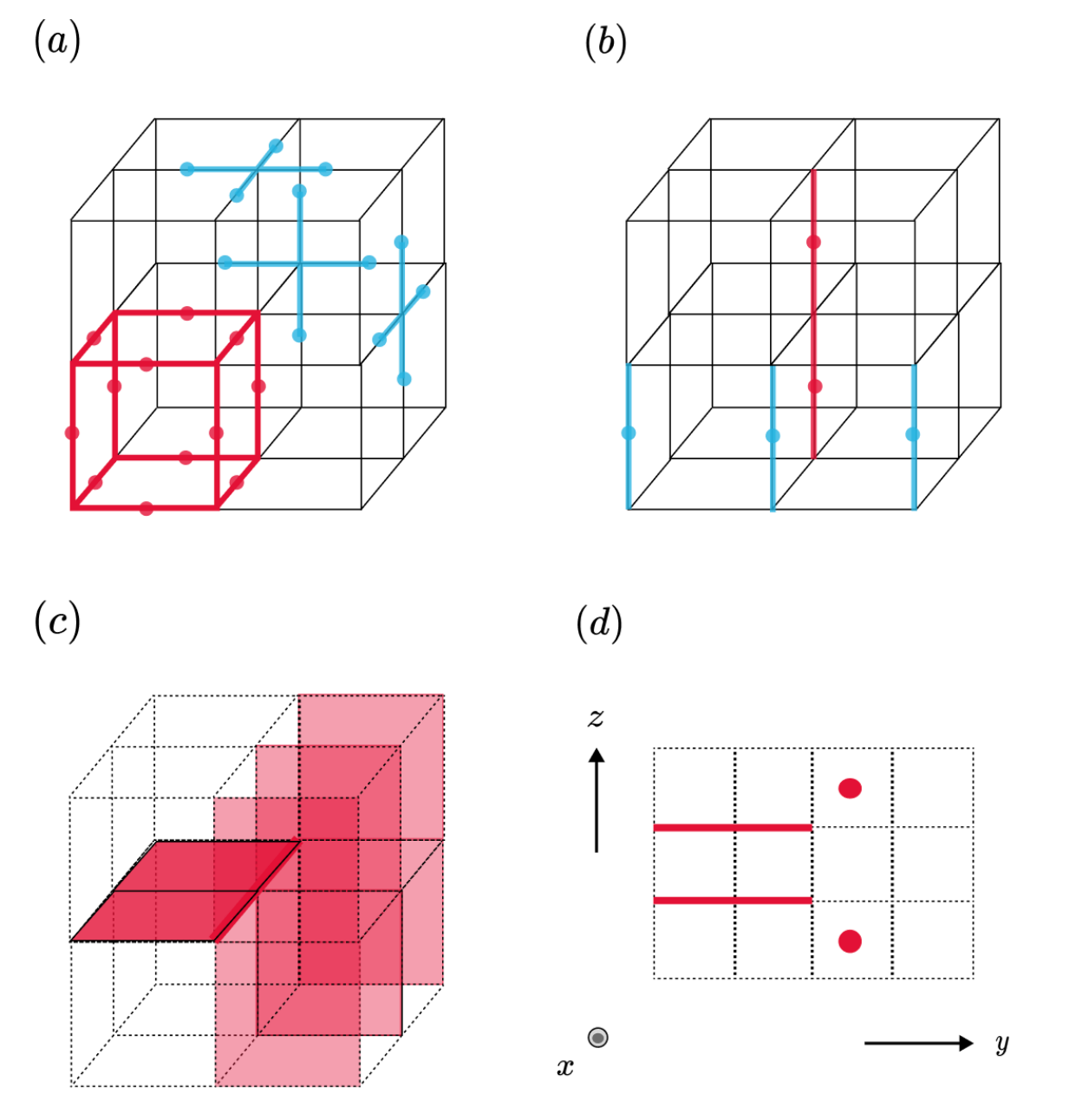}  
    \caption{\textbf{X-cube model.} {\bf (a) Stabilizers.} $X$-stabilizers (red) on cubes and three types of $Z$-stabilizers (blue) on vertices.  Each edge qubit is shared by four $X$-stabilizers or four $Z$-stabilizers. {\bf (b) Logical operators.}   Logical operators are one-dimensional strings. Logical $X$ operators that penetrate the vertices of some rectangle are not independent, since they are equal to the product of some cube stabilizers. Logical $Z$ operators in the $\hat{i}\hat{j}$ plane can be freely deformed in the $\epsilon_{ijk}\hat{k}$ direction by multiplying $Z$ type stabilizers.  {\bf (c,d) Domain walls in SM$_x$.} Applying the subsystem symmetry in the $y<0$ half of $\hat{x}\hat{y}$ planes creates a pair of domain wall excitation of plaquettes along non-trivial loops winding around $\mathbb{T}^3$ in the $\hat{x}$ direction. }
    \label{fig:X-cube}
\end{figure}

{\bf SM$_x$.} The model is defined on the dual cubic lattice, where the edges (qubits) become \emph{dual} faces and cubes ($X$-stabilizers) become \emph{dual} vertices. 
As four neighboring $X$ stabilizers share each edge qubit, SM$_x$ has 4 body interactions. The classical spins are placed on the centers of cubes, i.e., dual vertices, and there are 4-body interaction terms among spins within the same dual face. 
This model is called the plaquette Ising model~\cite{Mueller_2017}. 
To understand the global domain walls of subsystem symmetries, let us consider a specific example. Consider a subsystem symmetry acting on dual vertices of the $xy$-plane at a particular $z$-coordinate in the dual lattice ($\Pi^*$). Assume that it is truncated along $x$-direction as illustrated in \figref{fig:X-cube}(c). The truncated symmetry would change the signs of the terms the SM Hamiltonian near its boundary, which are characterized by the sequences of dual faces penetrated by $x$-direcitonal lines shown in \figref{fig:X-cube}(d). 
Therefore, $x$-directional domain wall is defined by a sequence of 4-body interactions with negative signs for dual faces penetrated by the $x$-direcitonal line as shown in \figref{fig:X-cube}(d). 
Although a pair of lines will be created at each truncated side, by stacking more truncated subsystem symmetries along $z$-direction, we can separate these domain walls. Therefore, a domain wall is given as a line nontrivially wrapping around $\mathbb{T}^3$ while penetrating dual faces. 
The creation of this domain wall exactly corresponds to the application of the logical $X$ operator in \figref{fig:X-cube}(a) along the $x$-direction.

{\bf SM$_z$.} The model is defined on the original cubic lattice. Note that for each vertex $v$, there are three different types of $Z$-stabilizers labeled by $\mu=x,y,z$. For each qubit on an edge $e\,{=}\,(v,v')$ along $\mu$-direction, it is shared by four $Z$-stabilizers at $v$ and $v'$ of types other than $\mu$.  
Therefore, SM$_z$ has four-body interaction terms with 3 types of classical spin variables $s^x, s^y, s^z$ living on the vertices. 
The $Z$-stabilizer constraint per vertex translates into $s^x_v s^y_v s^z_v = 1$. 
Therefore, we can choose $s^x$ and $s^y$ to be independent and identify $s^z = s^x s^y$. 
The resulting classical model is the random 3D anisotropically coupled Ashkin-Teller model (RACAT). This model has the subsystem symmetry that flips all the $s^{i}, s^{j}$ spins in a given $\hat{i}\hat{j}$ plane.

Without loss of generality, consider a subsystem symmetry acting on the $\hat{y}\hat{z}$ plane at $x=0$ restricted to the $z<0$ region; this will create a domain wall between $z<0$ and $z>0$ on this plane. 
The symmetry action restricted to $z<0$ would flip the sign of interactions terms of a type $\sim s^x_i s^y_i s^x_j s^y_j$ on $z$-direction edges. The domain wall along the nontrivial cycle in the $\hat{y}$ direction corresponds to the logical $Z$ operator illustrated in \figref{fig:X-cube}(b). 

\section{Conclusion}
\label{Discussion}
In this work, we exactly calculated the coherent information for general CSS codes under local incoherent Pauli errors and concluded that exact error correction is possible if and only if the ML decoder always succeeds in the thermodynamic limit. Thus, the fundamental threshold---independent of the decoding protocol--- is indeed saturated by the optimal decoder. We then considered a traditional mapping to random classical SM models and established a rigorous connection between the decoding transition of the quantum code and the phase transition in the underlying random classical SM model. We showed that the disorder-averaged free energy cost of inserting a domain wall in the random SM model is equivalent to the relative entropy $D(\rho||\rho^{\prime})$, and demonstrated $p_{\textrm{th}}^{\textrm{dec}} \leq p_{\textrm{th}}^{\textrm{coh}} \leq p_{\textrm{th}}^{\textrm{rel}}$ in the thermodynamic limit. This proves the Nishimori critical point to be a rigorous upper bound for the threshold of any decoder. We noted that $p_{\textrm{th}}^{\textrm{coh}}$ could be strictly smaller than $p_{\textrm{th}}^{\textrm{rel}}$ in principle, emphasizing the importance of carefully comparing the two thresholds. With the Nielsen-Schumacher condition providing clear-cut criteria for exact quantum error correction, our work concretely demonstrates how coherent information stands out as the key metric to precisely identify the decoding threshold. 

There are several topics left for future work. 
One direction is to further probe the associated random classical SM models for more exotic CSS codes, such as hyperbolic codes or qLDPC codes defined on \textit{expander graphs}. 
It could be interesting to study whether or not $p_{\textrm{th}}^{\textrm{coh}}=p_{\textrm{th}}^{\textrm{rel}}$ for such cases. 
Also, DIPT in cluster states---a well-known example for SPT phases---is expected to be understood similarly from the ''classical code'' point of view. This perspective may merit further investigation. Generalization to circuit-level noise models, or even non-stabilizer states might be interesting as well.

\begin{acknowledgments}
We are grateful to Sajant Anand, Yuto Ashida, Mio Murao, Hayata Yamazaki, and Satoshi Yoshida for fruitful discussions and comments. R.N. was supported by MEXT Quantum Leap Flagship Program (MEXT QLEAP) JPMXS0118069605, JPMXS0120351339. 
JYL is supported by the Simons Investigator Award and a faculty startup grant at the University of Illinois, Urbana-Champaign.
\end{acknowledgments}

\newpage
\bibliography{ref}

\newpage

\appendix
\section{Details of SGSOP}
\label{apA}
Here we describe the details of \textit{Symplectic Gram-Shmidt orthogonalization procedure} (SGSOP) \cite{Wilde_2009}. The procedure is as follows:
\begin{enumerate}
\item We start from $m$ Pauli tensors $g_{1}, g_{2}, \cdots g_{m}$, which are not necessarily commuting. Take $g_{1}$.\\
\item If $g_{1}$ commutes with all other operators $g_{2} \cdots g_{m}$, set it aside as "processed" operators and relable the remaining operators $g_{2} \cdots g_{m}$ as $g_{1} \cdots g_{m-1}$, respectively. Continue.\\
\item If $g_{1}$ anti-commutes with some operator $g_{j} (2\leq j \leq n)$, relable $g_{j}$ as $g_{2}$ and modify other operators as the following
\begin{equation}
    g_{i} \to g_{i}\cdot g_{1}^{f(g_{i}, g_{2})}\cdot g_{2}^{f(g_{i}, g_{1})}.
\end{equation}
Here, $f(g,h)=0$ if $g$ and $h$ commute and $f(g,h)=1$ if $g$ and $h$ anti-commute. $g_{i}$ after this procedure commutes with both $g_{1}$ and $g_{2}$, so set the pair $g_{1}$ and $g_{2}$ aside as "processed" operators.
\end{enumerate}
If we start from the normalizer of the CSS stabilizer code, we obtain the desired pairs of logical operators. Note that the initial normalizers are either $Z$-type or $X$-type by construction, so $X$-type operators are always mapped to $X$-type operators, and vice versa. Thus we can conclude that the logical operators of an arbitrary CSS stabilizer code can always be set as either $X$-type or $Z$-type. 

\section{Details of Kramers-Wannier dualness}
\label{app:KW}

Here we elaborate on the Kramers-Wannier dualness of the underlying SM models in the disorder-free limit. As mentioned in the main text, this duality is a natural consequence of the condition $H_xH_z^T=0$ and its transpose. Indeed, by performing the exact expansion of the partition function, one finds
\begin{align}\label{eq:hightemperature}
    \mathcal{Z}_x(\vb{o}, \vb{o}) 
    &=\frac{\sum_{\bm{s}_x}  \prod_l e^{\beta_x  \prod_{i \in \partial l}   s_{i}}}{2^{D_x} (2 \cosh \beta_x)^{n}} \nonumber\\
    &= \frac{1}{2^{D_x+n}}\sum_{\bm{s}_x}\prod_l \qty(1+\qty[\prod_{i \in \partial l}   s_{i}] \cdot \mathrm{tanh}\beta_x)\nonumber\\
    &= \frac{1}{2^{D_x+n}}\sum_{\bm{s}_x}\sum_{\bm{l}} \qty(\prod_{i \in \partial l}   s_{i} \cdot \mathrm{tanh}\beta_x)^{|\bm{l}|}
\end{align}
Note that the summation over $\bm{s}_x$ will vanish unless $\partial \bm{l}\,{=}\,0$. This is equivalent to the statement that $\bm{l} \in \ker H_x$. Therefore,
\begin{align}
    \mathcal{Z}_x(\vb{o}, \vb{o}) &= \frac{1}{2^{n-m_x}}\sum_{\bm{l}\in \mathrm{Ker} H_x}\qty(\mathrm{tanh}\beta_x)^{|\bm{l}|}\nonumber\\  
    &= \frac{1}{2^{n-m_x+D_z}} \sum_{\vb{k}_z} \sum_{\vb{s}_z} (\mathrm{tanh}\beta_x)^{|H_z^T\vb{s}_z+E^{\vb{k}_z}_{o}|}
\end{align}
where we have used the relation (\ref{eq:decomposition}) to decompose $\mathrm{Ker} H_x$ into $\mathrm{Im} H_z^T$ and the logical space. Notice that the final line includes summation over different boundary conditions (or put differently, different homology class). Assuming that the boundary effect is negligible, one gets
\begin{equation}\label{eq:approximate}
    \mathcal{Z}_x(\vb{o}, \vb{o}) \simeq \frac{1}{2^{m_z+D_z}} \sum_{\vb{s}_z} (\mathrm{tanh}\beta_x)^{|H_z^T\vb{s}_z|}.
\end{equation}
We can compare this result with the low-temperature expansion
\begin{equation}
    \mathcal{Z}_z(\vb{o}, \vb{o}) 
    =\frac{e^{n\beta_z}}{2^{D_z} (2 \cosh \beta_z)^{n}} \sum_{\bm{s}_z}   e^{-2\beta_z |H_z^T\vb{s}_z|}.
\end{equation}
Under the condition 
\begin{equation*}
e^{-2\beta_z}=\tanh \beta_x \Leftrightarrow \sinh 2\beta_x \sinh 2\beta_z =1,
\end{equation*}
we find the Kramers-Wannier duality relation
\begin{equation}
    \mathcal{Z}_x(\vb{o}, \vb{o}) = \frac{(1+ \tanh \beta_x)^{n}}{2^{m_z}}\mathcal{Z}_z(\vb{o}, \vb{o}).
\end{equation}
We remark that the approximation (\ref{eq:approximate}) may no longer be valid for codes with $k\to \infty$ \cite{Placke_2023}. Also, the duality derived here is for ferromagnetic configurations with no frustrations. 

\section{Relative entropy}
\label{app:relative}

Here we calculate the relative entropy for general CSS codes under local bit/phase-flip errors. Instead of starting from the maximally mixed logical subspace, we start from two different fixed logical sectors
\begin{align}
    &\rho_{0}=\ket{\bm{0}, \vb{k}_{0}}\bra{\bm{0}, \vb{k}_{0}}, \quad \bar{\sigma}_{i}^{z} \rho_{0} \bar{\sigma}_{i}^{z} = \rho_{0}\\
    &\rho_{0}^{\prime}=\ket{\bm{0}, \vb{k}_{0}^{\prime}}\bra{\bm{0}, \vb{k}_{0}^{\prime}}, \quad \bar{\sigma}_{i}^{z} \rho_{0}^{\prime} \bar{\sigma}_{i}^{z} = \rho_{0}^{\prime}.
\end{align}
Notice that the phase flip channel only changes the parity of $X$-type stabilizers; the logical $Z$ operators included in the error-chain only acts trivially on $\rho_{0}$ and $\rho_{0}^{\prime}$. Thus, the decohered density matrix after bit/phase-flip becomes 
\begin{align}\label{densitymatrix1}
    \rho&=\sum_{\vb{a}, \vb{b}, \vb{k}}P_{\vb{a}} P_{\vb{b}, \vb{k}-\vb{k}_{0}}\ket{\vb{a}, \vb{b}, \vb{k}}\bra{\vb{a}, \vb{b}, \vb{k}}\\\label{densitymatrix2}
    \rho^{\prime}&=\sum_{\vb{a}, \vb{b}, \vb{k}}P_{\vb{a}}P_{\vb{b}, \vb{k}-\vb{k}_{0}^{\prime}}\ket{\vb{a}, \vb{b},  \vb{k}}\bra{\vb{a}, \vb{b}, \vb{k}}.
\end{align}
For these two states, we calculate the relative entropy 
\begin{align}
    D(\rho||\rho^{\prime}) = \tr (\rho\mathrm{log}\rho)-\tr(\rho\mathrm{log}\rho^{\prime}).
\end{align}
It is infinite for the orthogonal initial pure states $\rho_{0}$ and $\rho_{0}^{\prime}$. The data-processing inequality ensures its monotonically decreasing behavior with respect to the error-rate. Considering these facts,  it is natural to expect that it remains infinite in the thermodynamic limit $n\to \infty$ so long as the error-rate is under the threshold. This corresponds to the indistinguishability of different logical sectors after decoherence. Plugging (\ref{densitymatrix1}), (\ref{densitymatrix2}) in, we get
\begin{align}
    D(\rho||\rho^{\prime}) &= \sum_{\vb{b}, \vb{k}}P_{\vb{b}, \vb{k}-\vb{k}_{0}}\mathrm{log}\frac{P_{\vb{b}, \vb{k}-\vb{k}_{0}}}{P_{\vb{b}, \vb{k}-\vb{k}_{0}^{\prime}}} \nonumber\\
    &= \langle \Delta F \rangle.
\end{align}
Since each $P_{\vb{b}, \vb{k}}$ can be expressed as the partition function of the associated random SM model, it can be concluded that the relative entropy is precisely the disorder-averaged free energy cost of inserting domain walls, which is usually used as the criterion to distinguish between decoding transitions in established arguments \cite{Dennis_2002}. 

\section{Optimal ML decoder} 
\label{app:optimal}

The maximum likelihood (ML) decoder is a deterministic decoder defined as follows; If the error chain $E$ incurs a parity change $(\vb{a}, \vb{b}, \vb{k}_x, \vb{k}_z)$, it has the conditional success probability
$$
P(\textrm{succ}|E) = 
\begin{dcases}
     1 \quad (\textrm{if}\,\, P_{\vb{a}, \vb{b}, \vb{k}_x, \vb{k}_z}=\max_{\vb{k}_{x}^{\prime}, \vb{k}_z^{\prime}}P_{\vb{a}, \vb{b}, \vb{k}_x^{\prime}, \vb{k}_z^{\prime}})\\
     0 \quad (\textrm{otherwise}).
 \end{dcases}
 $$
 In other words, the ML decoder corrects errors under the assumption that, given an error syndrome $(\vb{a}, \vb{b})$, the error chain $E$ comes from the most probable equivalence class $(\vb{a}, \vb{b}, \vb{k}_x, \vb{k}_z)$. Therefore, the total success probability of the ML decoder can be written as 
 \begin{align}
     P^{\textrm{suc.}} = \sum_{\vb{a}, \vb{b}} \max_{\vb{k}_x, \vb{k}_z}P_{\vb{a}, \vb{b}, \vb{k}_x, \vb{k}_z}.
 \end{align}
From the arguments in \cite{Kubica_2018},
\begin{equation*}\label{eq:success evaluation}
     2P^{\textrm{suc.}}-1 \leq \sum_{\vb{a}, \vb{b}, \vb{k}_x, \vb{k}_z}P_{\vb{a}, \vb{b}, \vb{k}_x, \vb{k}_z}\frac{P_{\vb{a}, \vb{b}, \vb{k}_x, \vb{k}_z}}{P_{\vb{a}, \vb{b}}} \leq P^{\textrm{suc.}}.
 \end{equation*}
 Thus, 
 \begin{equation}
     \sum_{\vb{a}, \vb{b}, \vb{k}_x, \vb{k}_z}P_{\vb{a}, \vb{b}, \vb{k}_x, \vb{k}_z}\frac{P_{\vb{a}, \vb{b}, \vb{k}_x, \vb{k}_z}}{P_{\vb{a}, \vb{b}}} =1 \Leftrightarrow P^{\textrm{suc.}}=1.
 \end{equation}

When this condition is satisfied, for any $\vb{k}' \neq 0$, we have
\begin{align}
     0 &\leq \sum_{\vb{a}, \vb{b}, \vb{k}_x, \vb{k}_z} P_{\vb{a}, \vb{b}, \vb{k}_x, \vb{k}_z} \frac{P_{\vb{a}, \vb{b}, \vb{k}_x, \vb{k}_z+\vb{k}^{\prime}}}{P_{\vb{a}, \vb{b}}} \nonumber\\
     &\leq \sum_{\vb{a}, \vb{b}, \vb{k}_x, \vb{k}_z} P_{\vb{a}, \vb{b}, \vb{k}_x, \vb{k}_z} \frac{P_{\vb{a}, \vb{b}}-P_{\vb{a}, \vb{b}, \vb{k}_x, \vb{k}_z}}{P_{\vb{a}, \vb{b}}} \nonumber \label{eq:cross evaluation}\\
    &= 1- \sum_{\vb{a}, \vb{b}, \vb{k}_x, \vb{k}_z}P_{\vb{a}, \vb{b}, \vb{k}_x, \vb{k}_z}\frac{P_{\vb{a}, \vb{b}, \vb{k}_x, \vb{k}_z}}{P_{\vb{a}, \vb{b}}} \to 0.
\end{align}
Leveraging on this relation, when $k \log 2\,{=}\,I_c$, we can show that the relative entropy for different logical sectors under bit/phase flip errors diverges
\begin{align}
     D(\rho||\rho^{\prime})=\langle \Delta F \rangle &= \sum_{\vb{b}, \vb{k}_z} P_{\vb{b}, \vb{k}_z} \mathrm{log} \frac{P_{\vb{b}, \vb{k}_z}}{P_{\vb{b}, \vb{k}_z+\vb{k}^{\prime}}} \nonumber\\
     &= \sum_{\vb{b}, \vb{k}_z} P_{\vb{b}, \vb{k}_z} \mathrm{log} \frac{P_{\vb{b}, \vb{k}_z}}{P_{\vb{b}}} \nonumber\\
     &- \sum_{\vb{b}, \vb{k}_z} P_{\vb{b}, \vb{k}_z} \mathrm{log} \frac{P_{\vb{b}, \vb{k}_z+\vb{k}^{\prime}}}{P_{\vb{b}}} \nonumber\\
     &= - \sum_{\vb{b}, \vb{k}_z} P_{\vb{b}, \vb{k}_z} \mathrm{log} \frac{P_{\vb{b}, \vb{k}_z+\vb{k}^{\prime}}}{P_{\vb{b}}} \nonumber\\
     & \geq  -\mathrm{log} \sum_{\vb{b}, \vb{k}_z} P_{\vb{b}, \vb{k}_z} \frac{P_{\vb{b}, \vb{k}_z+\vb{k}^{\prime}}}{P_{\vb{b}}} \rightarrow +\infty.
 \end{align}
See Appendix \ref{app:relative} for the derivation of the relative entropy.
As mentioned in the main text, we conclude from this result that
\begin{equation}
     p^{\mathrm{dec}}_{\textrm{th}} \leq p_{\textrm{th}}^{\textrm{coh}} \leq p_{\textrm{th}}^{\textrm{rel}}.
\end{equation}
Therefore, while the arguments based on free energy provide an upper bound on the optimal decoder's performance, the coherent information provides the tightest upper-bound. Notice that our exact calculation of coherent information eliminated the assumption $k<\infty $ in the original argument \cite{Kubica_2018}. This is a subtle but important point \cite{Placke_2023, Kovalev_2018}, since our result becomes applicable to codes with large numbers of logical qubits $k\to \infty$, such as hyperbolic surface codes \cite{Breuckmann_2016}, and good qLDPC codes~\cite{panteleev2022asymptotically,dinur_qldpc_decoder}. In such generic cases, the associated classical SM model may exhibit unconventional phase transitions along the Nishimori line~\cite{kovalev2014spin, Jiang_2019}, and a scenario is possible where the majority of the spins are ferromagnetic, while a constant fraction of spins remain paramagnetic. This situation could lead to $p_{\textrm{th}}^{\textrm{coh}} < p_{\textrm{th}}^{\textrm{rel}}$~\cite{lee2024exact}.

\section{General incoherent Pauli noise}
\label{Ynoise}

Here we consider local incoherent $Y$ noise as well. For CSS codes, the $Y$ noise is merely a correlated version of the $X$ and $Z$ noise.
\begin{equation}
    \sum_{E} P_{Y}(E) Y_{E} \rho_{0,Q} Y_{E} = \sum_{E} P_{Y}(E) X_{E} Z_{E} \rho_{0,Q} Z_{E} X_{E} 
\end{equation}
Thus, under the channel
\begin{align*}
    \mathcal{E}_{y,i}(\rho_{0,Q}) &= (1-p_{y})\rho_{0,Q} + p_{y}\sigma^{y}_{i}\rho_{0,Q}\sigma^{y}_{i}\\
    \mathcal{E}_{y} &= \prod_{i} \mathcal{E}_{y,i}, \quad \mathcal{E}= \mathcal{E}_{y}\circ \mathcal{E}_{z} \circ \mathcal{E}_{x}
\end{align*}
we get 
\begin{align}\label{y1}
    \rho_{Q}&=\mathcal{E}[\rho_{0,Q}]=\frac{1}{2^{k}}\sum_{\vb{k}}P_{\vb{a},\vb{b}}\ket{\vb{a}, \vb{b},\vb{k}}\bra{\vb{a}, \vb{b},\vb{k}}\\\label{y2}
    \rho_{QR}&=\sum_{\vb{a}, \vb{b}, \vb{k}_{x}, \vb{k}_{z}}P_{\vb{a}, \vb{b}, \vb{k}_{z}, \vb{k}_{z}}\qty[\prod_{i=1}^{k} \rho_{i}(\vb{k}_{z}, \vb{k}_{x})] \ket{\vb{a}, \vb{b}, \vb{k}} \bra{\vb{a}, \vb{b}, \vb{k}},
\end{align}
where $P_{\vb{a}, \vb{b}}$ corresponds to the probability that the parity for $X/Z$-type stabilizers become $\vb{a}\in \mathbb{F}_{2}^{m_{x}}, \vb{b}\in \mathbb{F}_{2}^{m_{z}}$ respectively. $P_{\vb{a}, \vb{b}, \vb{k}_{z}, \vb{k}_{x}}$ corresponds to the probability that the error-chain incurs the parity changes in logical $X$ and $Z$ operators by $\vb{k}_{x}, \vb{k}_{z}$ and in $X$ and $Z$ stabilizers by $\vb{a}, \vb{b}$. 
\begin{equation}\label{y3}
    P_{\vb{a}, \vb{b}} = \sum_{\vb{k}_{z}, \vb{k}_{x}} P_{\vb{a}, \vb{b}, \vb{k}_{z}, \vb{k}_{x}}. 
\end{equation}
Note that $\mathcal{E}$ includes the depolarizing channel 
\begin{equation}
    \mathcal{E}_{\mathrm{dep}}(\rho_{0}) = (1-P)\rho_{0}+\sum_{a=x,y,z}\frac{P}{3}\sigma_{i}^{a}\rho_{0}\sigma_{i}^{a}
\end{equation}
as its special case, with $p_{x}=p_{y}=p_{z}=p, P=3p(1-p)$. 
Plugging the above equations (\ref{y1}), (\ref{y2}), (\ref{y3}) in, we get
\begin{equation}\label{general}
    I_{c} = k\mathrm{log}2+\sum_{\vb{a}, \vb{b}, \vb{k}_{x}, \vb{k}_{z}}P_{\vb{a}, \vb{b}, \vb{k}_{x}, \vb{k}_{z}}  \log \frac{P_{\vb{a}, \vb{b}, \vb{k}_{x}, \vb{k}_{z}}}{\sum_{\vb{k}_{z}^{\prime}, \vb{k}_{x}^{\prime}}P_{\vb{a}, \vb{b}, \vb{k}_{x}^{\prime}, \vb{k}_{z}^{\prime}}}.
\end{equation}
In the above equation (\ref{general}), we cannot factorize $P_{\vb{a}, \vb{b}, \vb{k}_{z}, \vb{k}_{z}}$ due to the correlating effect of $Y$ noise. However, as elaborated in the main text, $P_{\vb{a}, \vb{b}, \vb{k}_{z}, \vb{k}_{z}}$ can be mapped to the partition function of some random SM model.
\nocite{*}

\end{document}